\documentclass[11pt]{article}
\pdfoutput=1
\usepackage{jheppub}
\usepackage{amsfonts}
\usepackage{amsmath,amssymb} 
\usepackage{graphicx} 
\usepackage[dvipsnames,x11names]{xcolor} 

\usepackage[nottoc,notlof,notlot]{tocbibind} 
\usepackage[titles]{tocloft} 
\usepackage{hyperref}
\usepackage{cleveref}
\usepackage{float}
\usepackage{subcaption}
\numberwithin{equation}{section} 
\usepackage{shuffle}
\usepackage{enumitem}

\begin{document}

\title{\textbf{Quantum chaos and the arrow of time}}

\author{Nilakash Sorokhaibam}\emailAdd{phy\_sns@tezu.ernet.in}

\affiliation{Department of Physics, Tezpur University, Tezpur, 784028, Assam, India.}

\abstract{We show that quantum chaotic many-body systems possess the thermodynamic arrow of time in the thermodynamic limit. Berry's conjecture in quantum chaotic systems and equivalence of ensembles imply the Kelvin statement of the second law of thermodynamics at leading order in perturbation theory. We verify this result using numerical calculations. We also show that this gives rise to new constraints on the off-diagonal terms in eigenstate thermalization hypothesis (ETH) statement. We call the new constraints collectively as ETH-monotonicity. These constraints arise because pure entropic consideration is not enough for the emergence of the thermodynamic arrow of time.}

\maketitle

\section{Introduction}
\label{sec:intro}
All physical systems around us become more and more disordered as time progresses unless an outside agent spends energy and record information for the upkeep of the system. This phenomenon has a beautiful statistical explanation. There simply are too many possible disordered states compared to the number of possible ordered states. So, the system ends up most likely in a disordered state \cite{Greiner1995}. Entropy is the quantity which measures the extent of disorder. In terms of entropy, this general observation gives rise to the second law of thermodynamics. It states that the entropy of an isolated system does not decrease. This is the Planck statement of the second law of thermodynamics \cite{planck1903treatise}. It gives rise to the thermodynamic arrow of time (called \emph{Time's arrow} in \cite{Eddington1927}). These results are familiar in the realm of classical physics. But our physical world is governed by quantum mechanics at microscopic level and so far a complete understanding of the quantum version of the second law is lacking. It is not clear if entropic consideration can fully explain the second law. In quantum mechanics, wavefunction plays a central role. So, denser regions of the energy spectrum are not necessarily readily available. Transition amplitudes to the denser region of the energy spectrum also play a critical role. For example, if the transition amplitudes to the energy eigenstates in the denser region is infinitesimally small then the wavefunction of the system would not transit towards this denser region of the energy spectrum. The central goal of this work is to study these transition amplitudes in quantum chaotic many-body systems.

The second law of thermodynamics has been stated as different statements all of which are considered to be equivalent. We will work with the Kelvin statement of the second law. It states that \emph{no process is possible whose sole result is the complete conversion of heat into work} \cite{Kardarbook2007,huang2008statistical}. In this paper, we will be working with the \emph{microscopic} theory so there will be no notion of heat. We will keep track of the energy of the system. So, the second law in the Kelvin form implies that it is impossible to convert energy from a single source into work without any other effect. In other word, if we perturb an otherwise closed system, the system cannot lose energy. If the system loses energy then the energy lost would have been completely converted into work. Examples of perturbing a system would be a momentary application of magnetic field in a spin system or shooting a pulse of laser at an electronic system.\footnote{We are not turning on the magnetic field indefinitely in which case the Hamiltonian of the system would be changed forever. We are also not considering cases in which the system is coupled to another system. An example of coupling two systems would be connecting the piston of an expanding gas to some other mechanical system like a dynamo through a crankshaft to extract electrical energy. If we let the gas expand without connecting the piston to the crankshaft, the total energy of the gas system would remain constant.} We have been considering positive temperature in our discussion. At negative temperature, the Kelvin statement implies that the system cannot gain energy when perturbed. At infinite temperature, there is no change in energy. In this paper, we will call the second law of thermodynamics in the form of the Kelvin statement as the arrow of time.

The relation between the Kelvin statement and the Planck statement can be easily verified using the thermodynamic relation
\begin{gather}
\Delta E=T \Delta S
\label{dETdS}
\end{gather}
where $T$, $\Delta E$ and $\Delta S$ are temperature, change in energy and  change in entropy. We have set the Planck constant $\hbar=1$ and the Boltzmann constant $k_B=1$. From the Kelvin statement, $\Delta E$ has the same sign as $T$ so $\Delta S$ is positive (or both $\Delta E$ and $\Delta S$ are equal to zero), which is the Planck statement $\Delta S \geq 0$.

In this work, we show that the arrow of time arises naturally in a isolated quantum chaotic system in the thermodynamic limit. We show that this result follows from an extension of Berry's conjecture \cite{Berry_1977,DAlessio:2015qtq} and equivalence of ensembles. Berry's conjecture states that expectation values of observables in a single energy eigenstate of a quantum chaotic system are same as their microcanonical averages. Quantum chaos in many-body systems and the related question of thermalization of these systems have been areas of intense research in the last few decades \cite{DAlessio:2015qtq}. Traditionally, quantum chaos is identified from the study of energy level statistics \cite{Bohigas:1983er} and of delocalization measures for the complexity of eigenstates \cite{Santos_PRE_2010}. Berry's conjecture deals with the diagonal elements of observables in the energy eigenstates. The off-diagonal matrix elements were first studied in detail in \cite{PhysRevA.34.591}. For many-body systems, a formal ansatz for the matrix elements is now well known as eigenstate thermalization hypothesis (ETH) \cite{PhysRevA.43.2046,Srednicki:1994mfb}. It is a comprehensive statement which explains how observables thermalize in a quantum system \cite{Rigol_2008nature,Polkovnikov:2010yn,DAlessio:2015qtq,Gogolin:2016hwy}. So, it is the route to statistical mechanics for quantum systems. ETH has been shown to hold true for physically sensible observables of quantum chaotic systems while it does not hold true for quantum integrable systems. So for the rest of the paper, we will assume that quantum chaos and ETH imply each other. Berry's conjecture is the diagonal part of ETH if we ignore the fluctuations. But ETH and thermalization do not immediately imply the second law of thermodynamics. We will discuss this matter in detail in section \ref{arrow}. So, the arrow of time gives rise to new constraints on the off-diagonal elements of ETH. The off-diagonal elements are the transition amplitudes between different energy eigenstates. The new constraint effectively says that, starting from an energy eigenstate, the \emph{bare} transition amplitudes without the entropic factor to nearby energy eigenstates on the denser side of the spectrum should be equal to or greater than the \emph{bare} transition amplitudes to the thinner side of the spectrum. The equality holds true in the thermodynamic limit when there is equivalence of ensembles. So, entropic consideration is not enough to give rise to the arrow of time. We will call these new constraints collectively as ETH-monotonicity.

The main idea of the paper is that Berry's conjecture should also hold true for change in energy when a chaotic system is perturbed. After all, change in energy after a system is perturbed is one of the most important quantity when studying a dynamical system. All the results in this paper will follow from this non-trivial but very reasonable assumption. We also verify this assumption using numerical calculations in the chaotic XXZ spin chain. At leading order in time-dependent perturbation theory, it can also be shown that a thermal state always possesses the arrow of time. So, Berry's conjecture implies that the quantum chaotic many-body systems in the thermodynamic limit possess the arrow of time. The logical flow of this work is as follows
\begin{gather*}
\text{Extension of Berry's conjecture to change in energy}\\
\text{+ Equivalence of ensembles}\\
\downarrow\\
\text{The thermodynamic arrow of time}\\
\downarrow\\
\text{New constraints on ETH}\
\end{gather*}

\begin{figure}
\begin{center}
\includegraphics[width=0.6\columnwidth]{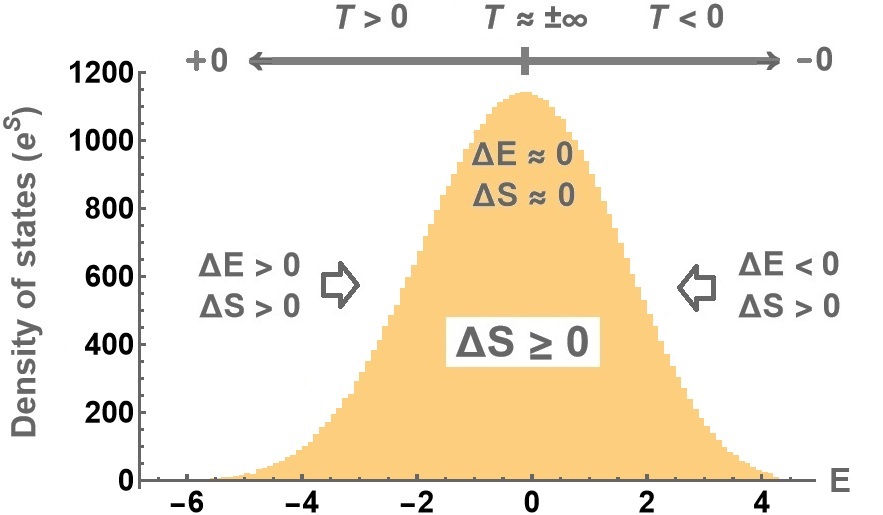}
\caption{\small The density of states ($e^S$) versus energy ($E$) plot for the chaotic XXZ spin chain. $T$ is effective temperature of the energy eigenstates. Perturbation of the system leads to a change in energy $\Delta E$ which agrees with the Kelvin statement of the second law of thermodynamics. Using the relation $\Delta E= T\Delta S$, it also implies the Planck statement $\Delta S\geq 0$.}
\label{summary_plot}
\end{center}
\end{figure}

It has previously been shown that the change in the \emph{diagonal} entropy is non-negative under doubly stochastic evolution starting from a \emph{passive} density matrix \cite{Polkovnikov_2011,DAlessio:2015qtq}. This follows from the convex nature of the function $x\log x$.  This result is true even for integrable systems. Thermal density matrix is an example of passive density matrix. But integrable systems do not thermalize after a finite perturbation. In the present work, we will be working with pure states which is the most appropriate approach for an isolated system. In the spirit of Berry's conjecture and ETH, we will be working with individual energy eigenstates.

Just as we cannot keep track of the positions and momenta of all gas molecules in the classical theory of gas, we cannot keep track of the complex coefficients of the exponentially large number of energy eigenstates of a quantum system. So in quantum mechanics, we keep track of only certain observables which can be easily measured or calculated. These can be intensive (local) operators or extensive (global) operators but they are few-body operators. If we observe that these observables thermalize then we say that the system thermalizes. ETH is a criterion for thermalization of such observables so that their long time expectation values can be described by quantum statistical mechanics. Inevitably, the quantum operators corresponding to these observables are hermitian and non-fermionic in nature.

Consider an operator $\mathcal{O}$ which corresponds to one such observable. ETH states that the matrix elements $\mathcal{O}_{mn}$ in energy eigenbasis are of the form
\begin{equation}
\langle m|\mathcal{O}|n\rangle\equiv \mathcal{O}_{mn}=\mathcal{O}(\bar{E})\delta_{mn}+e^{-S(\bar{E})/2} f(\bar{E},\omega)R_{mn}\
\label{ETH}
\end{equation}
where $|m\rangle$ and $|n\rangle$ are energy eigenstates with energies $E_m$ and $E_n$ respectively and $\bar{E}=(E_m+E_n)/2$, $\omega=E_m-E_n$. $S(\bar{E})$ is the entropy at energy $\bar{E}$. $\mathcal{O}(\bar{E})$ is equal to the expectation value of $\mathcal{O}$ in the microcanonical ensemble at energy $\bar{E}$ or other ensembles by the equivalence of ensembles. $\mathcal{O}(\bar{E})$ and $f(\bar{E},\omega)$ are smooth functions of their arguments. $f(\bar{E},\omega)$ is taken to be a real and positive function. $R_{mn}$ are pseudo-random variables with zero mean and unit variance \cite{Foini:2018sdb}. $\mathcal{O}$ is hermitian and hence
\begin{equation}
\mathcal{O}_{nm}=\mathcal{O}^*_{mn}, \; R_{nm}=R^*_{mn}, \; f(\bar{E},-\omega)=f(\bar{E},\omega)
\end{equation}

The $\omega$-dependence of the $f$-function has been a subject of study in many works \cite{Khatami_2013,Srednicki_1999,DAlessio:2015qtq,Dymarsky:2018ccu,Murthy:2019fgs,Wang:2021mtp}. At large $|\omega|$, the $f$-function falls exponentially faster than $e^{-|\beta\omega|/4}$ where $\beta=1/T$ is the inverse effective temperature \cite{Srednicki_1999,DAlessio:2015qtq,Dymarsky:2018ccu,Murthy:2019fgs}. Moreover, for small $\omega \sim 1/L^2$, where $L$ is the system size, $f$-function can be assumed to be a constant function as in case of random matrix theory (RMT) \cite{Khatami_2013,DAlessio:2015qtq,Dymarsky:2018ccu,Wang:2021mtp}.

In this work, we will study the $\bar{E}$-dependence of the $f$-function which is also closely related with the arrow of time. To the best of our knowledge, the $\bar{E}$-dependence of the $f$-function has not been studied anywhere. We will use the shorter notation $f(\bar{E})$ for $f(\bar{E},\omega)$ at a constant $\omega$ when the $\omega$-dependence is not important. We will also work with $f(\bar{E})$ as a function of the effective temperature $T(\bar{E})$ or the effective inverse temperature $\beta(\bar{E})$ or the entropy $S(\bar{E})$. With a slight abuse of notation, we will represent these functions as $f(T)$ or $f(\beta)$ or $f(S)$. But what these functions really mean are
\begin{gather}
f(T)=\tilde{f}(T(\bar{E}))=f(T^{-1}(T(\bar{E}))), \qquad f(\beta)=\breve{f}(\beta(\bar{E}))=f(\beta^{-1}(\beta(\bar{E}))),\nonumber\\ f(S)=\check{f}(S(\bar{E}))=f(S^{-1}(S(\bar{E}))).\
\end{gather}
$S^{-1}$ function can be multivalued. We have to choose the appropriate value by crosschecking with values of $T$ or $\beta$. We found that
\begin{center}
\emph{$f$-function is a monotonically increasing function of the magnitude of the temperature $|T(\bar{E})|$ or the entropy $S(\bar{E})$. Moreover, $f(\bar{E})$ flattens as $L^{1}$ as the system size $L$ increases,}
\begin{gather}
\frac{f'(\bar{E})}{f(\bar{E})}\;\xrightarrow{L\to\infty} \;0, \qquad L\;\frac{f'(\bar{E})}{f(\bar{E})}\; = \; \text{a constant.}
\label{ETHmono}
\end{gather}
\end{center}
This is our technical result. ETH-monotonicity is concerned with this $\bar{E}$-dependence of $f$-function. In summary, our results are
\begin{enumerate}[label=R{{\arabic*}}.]
\item We show that Berry's conjecture in quantum chaotic systems and equivalence of ensembles imply that quantum chaotic many-body systems possess the arrow of time.
\item $f$-function is a constant function of $\bar{E}$ in the thermodynamic limit.
\item $f$-function is a monotonically increasing function of the magnitude of the temperature $|T(\bar{E})|$ or the entropy $S(\bar{E})$.
\item $f(\bar{E})$ flattens as $L^{1}$ as the system size $L$ increases.
\end{enumerate}
R1 and R2 are analytic results. R3 and R4 are numerical results. We also provide some analytic motivations for R3. ETH-monotonicity comprises the results R2, R3 and R4. $\bar{E}$ is an extensive quantity but we will see that the flattening of $f(\bar{E})$ is not merely stretching of this function as the system size increases. Temperature $T$ is an intensive quantity so the monotonically increasing property of $f(T)$ will hold true even in the thermodynamic limit. Our numerical results are performed for finite systems with finite number of degrees of freedom. But we expect that our results will hold true even for continuum systems like quantum field theories.

We will be mostly working with pure states. The temperature $T$ of a pure state is the effective temperature calculated using equivalence of ensembles. Basically we calculate the temperature of the canonical ensemble having the same energy as the pure state. Correspondingly, the entropy (of even a single eigenstate) is the microcanonical entropy. For a clear perspective, the difference between canonical and microcanonical expectation values can be seen in \cite{Rigol_2008nature}. In Appendix \ref{cmccomp_app}, we will also find that the canonical entropy is always greater than the microcanonical entropy. The energy window of microcanonical ensemble $dE$ is usually considered to be an arbitrary constant as long as it does not grow extensively with system size. For this work, we need a precise definition of the energy window. We found a natural definition of the energy window by a comparative study of the microcanonical entropy and canonical entropy. Our choice of the energy window is such that the ratio of the canonical entropy and microcanonical entropy remains constant throughout the energy spectrum. The details can be found in Appendix \ref{cmccomp_app}. We also found that this optimum energy window does not grow extensively with the system size. In fact, for the three different system sizes under consideration we found that the optimum energy window to be a constant and equal to $dE=0.8$. In Appendix \ref{cmccomp_app}, we also compare our choice of the energy window with the energy window used in \cite{Rigol_2008nature}. We believe that our new definition of the energy window is more natural since it arises from the comparison of entropies.

In section \ref{arrow}, we will touch upon the related topic of thermalization and we will argue that thermalization is different from the arrow of time. Section \ref{set-up} describes the physical set-up and the main results. Section \ref{1stpass} shows that Berry's conjecture and the equivalence of ensembles imply the arrow of time. Section \ref{2ndpass} studies the details which give rise to our technical result - ETH-monotonicity. Section \ref{higher_sec} covers higher order terms in the perturbation theory. Numerical results of large perturbations are presented in section \ref{lperturb}. Section \ref{concl} is conclusions. The appendix consists of sections with lengthy mathematical derivations and other supplementary materials. By perturbation, we mean disturbing the system, whereas perturbation theory means the time-dependent perturbation theory of quantum mechanics.

\section{Thermalization does not imply the arrow of time}
\label{arrow}
The most important consequence of ETH is thermalization of non-equilibrium excited states which are typical \cite{Srednicki_1999}. Typical states are states with sub-extensive energy variance.\footnote{Note that this is the definition of a typical state in ETH works and it has no relation with other forms of quantum typicality studied in quantum statistical mechanics and quantum thermodynamics.} If an observable satisfy the ETH statement then its expectation value in a typical state relaxes to the microcanonical average specified by the fixed conserved charges in the long time limit. Fluctuations away from the microcanonical average can be interpreted as thermal fluctuations \cite{Srednicki:1995pt}, as long as we do not wait long enough to observe Poincare recurrence. This scenario is most popularly studied as quantum quenches where the non-equilibrium state is prepared by changing the Hamiltonian of the system \cite{Polkovnikov:2010yn,Gogolin:2016hwy}. The initial Hamiltonian and the final Hamiltonian are different.

One subtlety when preparing states using quantum quenches is that even integrable systems equilibrate to generalized Gibbs ensembles (GGEs) in the large system size limit, if the initial state has no long range correlation. Equilibration of quantum integrable systems after quantum quenches has been a topic of great interest in the last two decades \cite{Rigol_2007,Mandal:2015jla,Mandal:2015kxi,Polkovnikov:2010yn,Gogolin:2016hwy,Banerjee:2019ilw}. The non-equilibrium states are generalized typical states taking into account all the other conserved charges of the integrable system. This result follows from central limit theorem and the local nature of the Hamiltonians \cite{Rigol_2008nature,DAlessio:2015qtq}. The difference between chaotic systems and integrable systems arises when we consider finite perturbations for a finite duration of time (also called \emph{critical to critical quench} in \cite{Mandal:2015kxi} and \emph{bump quenches} in \cite{Bhattacharya:2018fkq}). The initial Hamiltonian and the final Hamiltonian are same. A chaotic system still thermalizes after such finite perturbations  \cite{Bhattacharya:2018fkq} while an integrable system cannot be even described by a GGE in long time limit \cite{Mandal:2015kxi}. The temperature and other chemical potentials become imaginary. So in this manner, finite perturbations or bump quenches have a potential to bring out the stark contrast between non-equilibrium dynamics of quantum chaotic systems and quantum integrable systems.

The apparent one-way evolution of a thermalization process, even when the microscopic dynamics is time-reversal invariant, has been considered to be an ``arrow of time" \cite{Srednicki_1999}. But we note that thermalization is different from our definition of the thermodynamic arrow of time. First of all, just because a system thermalization does not imply that the change in energy $\Delta E$ has the same sign as the temperature $T$ which is what the Kelvin statement of the second law of thermodynamics implies. Indeed we expect that the system would thermalize back to a new thermal equilibrium after the perturbation has stopped, but this thermalization does not imply that $\Delta E$ will strictly be either positive or negative. In other word, although thermalization and the arrow of time under consideration are physical phenomena observed in a chaotic system, they do not imply each other.

Another aspect of the thermalization process in systems (with discrete energy spectrum) is that one can observe Poincare recurrence if one waits long enough. The system will return arbitrarily close to its initial state if one waits long enough, although the associated timescale called Poincare recurrence time grows exponentially as the system size. But the fact that a system gains energy $\Delta E > 0$, for example, is permanent since the system is isolated except for the finite-time perturbation.

So, we want to emphasize that thermalization is not directly related with the arrow of time under consideration. As a matter of fact, we are not concerned with the non-equilibrium dynamics after the perturbation to the system has stopped. Nevertheless, we expect that our system would thermalize since it is a chaotic system and its observables satisfy ETH.

\section{Set-up and results}
\label{set-up}
Consider a system with the time-independent Hamiltonian $H_0$ and initially in a state $|\psi\rangle$. The system is perturbed by a time-dependent term $\lambda(t)\mathcal{O}$ where $\mathcal{O}$ is a non-fermionic, hermitian operator. $\mathcal{O}$ could be either an intensive operator or an extensive operator. The total Hamiltonian is
\begin{equation}
H(t)=H_0+\lambda(t)\mathcal{O}
\label{Ht}
\end{equation}
The perturbation is for a finite duration of time, say, from time $-d$ to $d$. The source $\lambda(t)$ is real and can be non-zero only for $-d<t<d$. This perturbing term is not necessarily small but it also cannot be arbitrarily big. The change in energy of the system is
\begin{equation}
\Delta E=\langle \psi(t_f)|H_0|\psi(t_f)\rangle-\langle \psi(t_i)|H_0|\psi(t_i)\rangle\
\label{dE1}
\end{equation}
where $t_i<-d$ and $d<t_f$. Our main goal is to show that $\Delta E$ is positive (negative, zero) if the initial temperature is positive (negative, infinite) when $H_0$ is chaotic and $\mathcal{O}$ satisfies ETH. This would imply that $\Delta S \geq 0$.

For now, the initial state is taken to be an energy eigenstate $|n\rangle$. We will consider other pure states later. Using the Dyson series expansion of the time evolution operators, we can perform an expansion of $\Delta E$ in powers of $\lambda(t)$. We will concentrate on the leading term which is given by
\begin{equation}
\Delta E_n=\sum_m \Delta E_{nm}=\sum_m (E_m-E_n)|\tilde{\lambda}(E_m-E_n)|^2|\mathcal{O}_{mn}|^2+O(\lambda^3)\
\label{dEl2}
\end{equation}
where $\tilde{\lambda}(\omega)=\tilde{\lambda}(-\omega)^*$ is the Fourier transform of $\lambda(t)$. The derivation of this expression can be found in Appendix \ref{dEn_app}. Whether $\Delta E$ is positive, negative or zero is not a priori clear. It depends on the $\mathcal{O}_{mn}$ elements. In this set-up, the ETH statement is a statement about the transition amplitude $|\mathcal{O}_{mn}|^2$ from one energy eigenstate $|n\rangle$ to another energy eigenstate $|m\rangle$.

For numerical calculations, we consider the XXZ spin chains with open boundary condition. With a large next-to-nearest neighbour interaction, the system is chaotic \cite{Gubin_2012}. We consider a system of size $L$ with the number of up-spins $N$. So, total magnetization is $2N-L$. The numerical methods involved are described in Appendix \ref{numres_app}. The system is perturbed using the global spin-current operator or the kinetic energy operator (the hopping term). It has been shown that these operators satisfy the ETH statement in the chaotic XXZ spin chain \cite{Steinigeweg_2013}. Both the operators preserve the total magnetization.

\begin{gather}
H_0=\sum_{i=1}^{L-1}\left[J_{xy}\left(S_i^x S_{i+1}^x+S_i^y S_{i+1}^y\right)+J_zS_i^z S_{i+1}^z\right]+\sum_{i=1}^{L-2}J'_zS_i^z S_{i+2}^z\\
\mathcal{O}_{SC}=\sum_{i=1}^{L-1}J_{xy}\left(S_i^xS_{i+1}^y-S_i^yS_{i+1}^x\right), \qquad
\mathcal{O}_{KE}=\sum_{i=1}^{L-1}J_{xy}\left(S_i^x S_{i+1}^x+S_i^y S_{i+1}^y\right).\
\end{gather}
We take $J_{xy}=1$ and $J_z=J'_z=0.5$. We work with three different system sizes - $L=14, N=7$; $L=16, N=8$; and $L=18, N=9$. So, we are only considering a subspace with zero magnetization out of the full state space which is $2^L$ dimensional. The corresponding dimensions of the state (sub) spaces are $\binom{14}{7}=3432$; $\binom{16}{8}=12870$; and $\binom{18}{9}=48620$. $L/N$ ratio is kept constant so that the large system size limit is the correct thermodynamic limit.

We will also study the ETH-monotonic behaviour of a single site spin operator at the middle of the spin chain $S^z_{L/2}$. The results for this operator will be presented in Appendix \ref{numres_app}. We do not specifically consider perturbing the system with this operator because it couples state subspaces with different magnetization.

\subsection{First pass: the arrow of time}
\label{1stpass}
The main idea that we will use in this subsection is that Berry's conjecture should also hold true for change in energy when a chaotic system is perturbed. So the change in energy after a perturbation starting from an energy eigenstate should be same as the change in energy starting from a microcanonical ensemble. Otherwise, one would be able to differentiate between an energy eigenstate and the corresponding microcanonical state by performing a small perturbation experiment and measuring the change in energy. Moreover, the equivalence of ensembles would imply that the change in energy starting from the energy eigenstate should be same as change in energy starting from a canonical ensemble in the thermodynamic limit. On the other hand, this proposal would not hold true for integrable systems.

The change in energy starting from a canonical ensemble with the density matrix $e^{-\beta H_0}$ is given by
\begin{gather}
\label{LRTresult}
\Delta E_{\beta}=\frac{1}{2}\int_{-\infty}^{\infty} d\omega \; \omega \; |\lambda(\omega)|^2\, A_{\beta}(\omega),\\
A_{\beta}(\omega)=\frac{1}{Z}\left(e^{\beta \omega}-1\right)\sum_{n,m}  e^{-\beta E_m} |\mathcal{O}_{nm}|^2\delta(\omega-(E_m-E_n)),\
\label{spectralfndef}
\end{gather}
where $Z$ is the partition function and $\tilde{\lambda}(\omega)=\tilde{\lambda}^*(-\omega)$ is the Fourier transform of $\lambda(t)$. What is remarkable is that the above expression of $\Delta E_{\beta}$ is positive (negative or zero) if the initial temperature is positive (negative or infinite) which is precisely the statement of the second law of thermodynamics in the Kelvin form. This is a mathematical identity. The above result is derived using two different techniques in Appendix \ref{dEtherm_app}. $A_{\beta}(\omega)$ is the spectral function. It is also equal to $-2\,\text{Im}\;\tilde{G}_R(\omega)$ where $\tilde{G}_R(\omega)$ is the Fourier transform of the retarded Green function $G_R(t,t')=-i\theta(t-t')\langle[\mathcal{O}(t),\mathcal{O}(t')]\rangle$. So Berry's conjecture and equivalence of ensemble implies that the quantum chaotic many body systems possess the arrow of time in the thermodynamic limit.

\begin{figure}[h]
\centering
\includegraphics[width=.8\columnwidth]{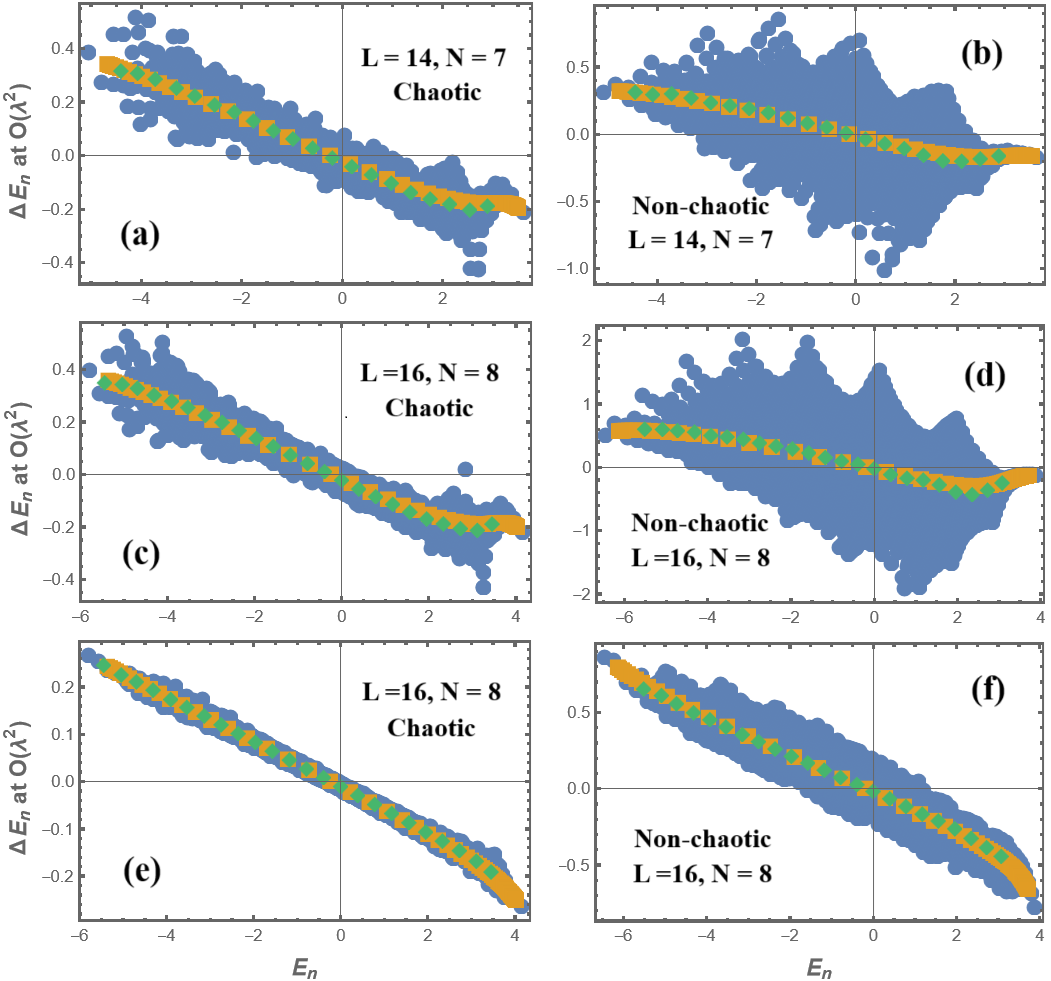}
\caption{\small{The leading term of the change in energy $\Delta E_n$ for the chaotic spin chains (left panels) and non-chaotic spin chains (right panels) after a unit delta function perturbation $\lambda(t)=\delta(t)$ with spin current operator $\mathcal{O_{SC}}$ (panels a, b, c and d), and with the kinetic energy term $\mathcal{O}_{KE}$ (panels e and f). The blue dots are for the different energy eigenstates of energy $E_n$ as the initial state, the yellow plots are for canonical ensembles and the green dots are for microcanonical ensembles. We have considered spin chain of system size $L=14, N=7$ (panel a and b), and $L=16, N=8$ (panel c, d, e and f). The dimension of the state space are $\binom{14}{7}=3432$ and $\binom{16}{8}=12870$. So, there are 12870 blue dots in each of the panels c, d, e and f.}}
\label{6panels_l2}
\end{figure}

Figure \ref{6panels_l2} are the plots of the leading term in $\Delta E_n$ after a unit delta function perturbation $\lambda(t)=\delta(t)$ in each of the energy eigenstates $|n\rangle$. In the chaotic system, the leading terms starting from different energy eigenstates match the thermal values considering both canonical and microcanonical ensembles. As one expects, the matching is better for larger systems. In case of the integrable system, the leading term starting from different energy eigenstates wildly vary compared to the thermal values.

Now let us consider a more general state. Consider a typical state $|\Psi\rangle=\sum_n c_n|n\rangle$. The $c_n$ coefficients are non-zero for a small energy window similar in size to the microcanonical energy window ($dE$ in Appendix \ref{cmccomp_app}). Now the change in energy after the system is perturbed will have diagonal terms and off-diagonal terms in the $|n\rangle$ basis. We will not be concerned with the various energy scales associated with typical states \cite{Srednicki_1999}. Our main motive is to show that the off-diagonal terms average out to zero. The change in energy starting from $|\Psi\rangle$ is
\begin{gather}
\label{typical_state}
\Delta E_{\Psi}=\sum_n |c_n|^2\Delta E_n+\sum_{n,n'\neq n} c^*_nc_{n'}\sum_m\Delta E_{nn'm},\\
\sum_m\Delta E_{nn'm}=\sum_m \mathcal{O}_{nm}\mathcal{O}_{mn'}\left[\int_{-\infty}^{\infty}dt_1dt'_1\lambda(t_1)\lambda(t'_1)\,E_m e^{it_1(E_n-E_m)+it'_1(E_m-E_{n'})}\right.\hspace{3cm}\nonumber\\
\left.\hspace{3cm}-\int_{-\infty}^{\infty}dt_1\int_{-\infty}^{t_1}dt_2\,\lambda(t_1)\lambda(t_2)\left\{E_ne^{it_2(E_n-E_m)+it_1(E_m-E_{n'})}+(t_1 \leftrightarrow t_2, n \leftrightarrow n')\right\}\right],\nonumber\
\end{gather}
where sum over $m$ is over the entire state space. $R_{mn}$ has zero mean so the off-diagonal terms average out to zero, so the second sum in (\ref{typical_state}) is zero. It is worth noting that this will not happen in integrable systems. The (pseudo-) randomness in $R_{mn}$ is at the level of the energy gap between adjacent energy levels which is exponentially small in system size so all the other factors can be effectively taken to be constants compared to the variation of $R_{mn}$. In fact, the energy gap between adjacent energy levels is the smallest energy scale in the system. The diagonal terms give the microcanonical average which is the expected result. Averaging of $R_{mn}$ is routinely used in all ETH works, see for example in \cite{Srednicki_1999,Murthy:2019fgs}. We will use it again in the following subsection. We can also consider perturbation with multiple operators satisfying ETH.
\begin{equation}
H(t)=H_0+\sum_i\lambda_i(t)\mathcal{O}_i
\label{Ht_multi}
\end{equation}
In this case also, the contribution to the change in energy from the cross-terms $\mathcal{O}_i\mathcal{O}_{i'}$, with $i\neq i'$, will be zero because of $R_{i,mn}$ and $R_{i',mn}$.

\subsection{Second pass: new constraints on ETH statement}
\label{2ndpass}
In this subsection, we will do a detailed comparison of (\ref{dEl2}) and (\ref{LRTresult}) which gives the new constraints that we are calling ETH-monotonicity. We use the ETH statement (\ref{ETH}) in (\ref{LRTresult}). $|R_{mn}|^2$ is replaced by its average value which is equal to one. We substitute $\sum_{m}\to\int_{-\infty}^{\infty}dE_m e^{S(E_m)}$ and $\sum_{n}\to\int_{-\infty}^{\infty}dE_n e^{S(E_n)}$ and convert the integration variables into $\omega'=E_m-E_n$ and $\bar{E}=(E_m+E_n)/2$. (\ref{spectralfndef}) becomes
\begin{eqnarray}
A_{\beta}(\omega)&=&(e^{\beta\omega}-1)\int^{\infty}_{-\infty} d\omega' d\bar{E}\, e^{S(\bar{E}+\omega'/2)+S(\bar{E}-\omega'/2)-S(\bar{E})}e^{-\beta\bar{E}-\beta\omega'/2}f(\bar{E},\omega)^2\delta(\omega-\omega')\nonumber\\
&=&(e^{\beta\omega/2}-e^{-\beta\omega/2})\int^{\infty}_{-\infty} d\bar{E}\, e^{S(\bar{E})-\beta\bar{E}}f(\bar{E},\omega)^2\nonumber\\
&=&2\sinh(\beta\omega/2)f(\bar{E}_{\beta},\omega)^2\
\end{eqnarray}
The $\bar{E}$-integral fixes $\bar{E}$ to be $\bar{E}_{\beta}$, the energy corresponding to the inverse temperature $\beta$, while going from the second line to the third line. In the Taylor expansion of $S(\bar{E}+\omega'/2)+S(\bar{E}-\omega'/2)-S(\bar{E})$ about $\bar{E}$, the higher order $\omega$-terms are suppressed by powers of the system size. For example, the second order derivative is suppressed by the total heat capacity of the system \cite{Murthy:2019fgs}. So, the change in energy starting from the canonical ensemble is
\begin{equation}
\Delta E_{\beta}=\int_{-\infty}^{\infty}d\omega \, \omega|\lambda(\omega)|^2 \sinh(\beta\omega/2)f(\bar{E}_{\beta},\omega)^2\
\label{LRTfinal}
\end{equation}
Now it is clear that $\Delta E_{\beta}$ have the same sign as $\beta$ and is zero when $\beta=0$. $|\lambda(\omega)|$ and $f(\bar{E},\omega)$ are even functions of $\omega$.

Substituting $\sum_{E_m}\to\int_{-\infty}^{\infty}dE_m e^{S(E_m)}$ and changing the integration variable to $\omega=E_m-E_n$, (\ref{dEl2}) becomes
\begin{equation}
\Delta E_n=\int_{-\infty}^{\infty}d\omega\, \omega |\lambda(\omega)|^2\,e^{S(E_n+\omega)-S(E_n+\omega/2)} f(E_n+\omega/2,\omega)^2
\label{dEexactO2}
\end{equation}
In a finite system with finite dimensional state space, the entropic factor $e^{S(E_n+\omega)-S(E_n+\omega/2)}$ will have a peak at $\omega=\omega_0$ which has the same sign as $\beta$. We study this entropic factor in detail in Appendix \ref{entfac_app}. We show that the position of the peak of this entropic factor moves away to $\omega\to\pm\infty$ in the thermodynamic limit. For such large systems, the convergence of the above integral is ensured by the $\omega$-dependence of $f(\bar{E},\omega)$ which falls exponentially faster than $e^{-|\beta\omega|/4}$ for large $\omega$. So for large systems, we can perform Taylor expansion of $S(E_n+\omega)-S(E_n+\omega/2)$ about $E_n$. Again, the higher order derivatives are suppressed by powers of the system size. This reduces the change in energy to
\begin{equation}
\Delta E_n=\int_{-\infty}^{\infty}d\omega\, \omega |\lambda(\omega)|^2\,e^{\beta\omega/2} f(\bar{E}_{\beta}+\omega/2,\omega)^2
\label{microDE}
\end{equation}
where $\beta$ is the inverse temperature of $E_n$. Note also that we have replace $E_n$ with $\bar{E}_{\beta}$ in the expression above so that no confusion arises when we compare this expression with (\ref{LRTfinal}). The $e^{\beta\omega/2}$ factor is the entropic factor in support of the arrow of time. It comes from the competition of the entropic factor from the sum over the state space and the entropic factor of the off-diagonal term in the ETH statement. So, $f(\bar{E}_{\beta}+\omega/2,\omega)^2$ acts like a \emph{bare} transition amplitude in a quantum chaotic system. Note that the real transition amplitude between energy eigenstates $|n\rangle$ and $|m\rangle$ is simply $|\mathcal{O}_{mn}|^2$. An interesting observation is that $f(\bar{E})$ should be a flat function at $\beta=0$ otherwise $\Delta E_n$ will be non-zero which is not expected from the Kelvin statement.

The difference between (\ref{LRTfinal}) and (\ref{microDE}) is the $\omega/2$ in the first argument of the $f$-function in (\ref{LRTfinal}). So, if we perform a Taylor expansion of $f(\bar{E}_{\beta}+\omega/2,\omega)$ about the point $(\bar{E}_{\beta},\omega)$ and consider only the first order term, then we get
\begin{equation}
\Delta E_n=\int_{-\infty}^{\infty}d\omega \, \omega|\lambda(\omega)|^2 e^{\beta\omega/2}f(\bar{E}_{\beta},\omega)^2\left[1+\left.\left(\frac{\omega}{f(\bar{E},\omega)}\frac{\partial f(\bar{E},\omega)}{\partial\bar{E}}\right)\right|_{\bar{E}=\bar{E}_{\beta}}\right]\
\end{equation}
$|\tilde{\lambda}(\omega)|$ and $f(\bar{E},\omega)$ are even functions of $\omega$,
so the first term inside the square bracket gives us the thermal value (\ref{LRTfinal}). Then, the second term inside the square bracket should vanish in the thermodynamic limit. It is a competition with a number, namely one, not a function of $\bar{E}$. It means that $f(\bar{E},\omega)$ flattens, 
\begin{equation}
\frac{f'(\bar{E})}{f(\bar{E})}\quad \xrightarrow[\text{limit}]{\text{Thermodynamic}} \quad 0.
\label{fflat}
\end{equation}
So, the \emph{bare} transition amplitude is constrained by the arrow of time in a quantum chaotic system. We will present here numerical results for the spin current operator (an extensive operator). Numerical results for the kinetic energy operator (another extensive operator) and the single site spin operator (an intensive operator) can be found in Appendix \ref{numres_app}. Figure \ref{fEb_Eb_innNNNL18_omega0125_plot} are plots of the $f$-function as a function of $\bar{E}$ for different system sizes. The flattening of $f(\bar{E})$ is not clear from these plots.

\begin{figure}
\begin{center}
\includegraphics[width=.8\linewidth]{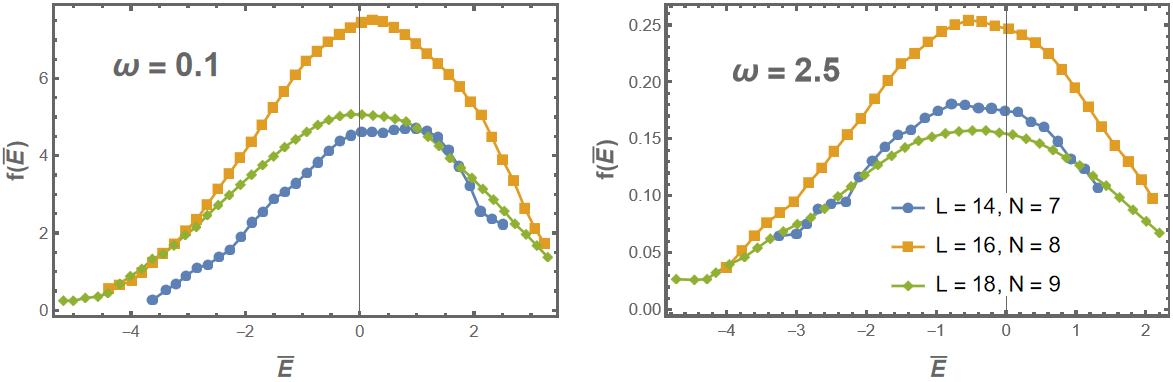}
\caption{\small{$f(\bar{E})$ plots with fixed $\omega$'s for the spin current operator with different system sizes $L=14, N=7$, $L=16, N=8$, and $L=18, N=9$.}}
\label{fEb_Eb_innNNNL18_omega0125_plot}
\end{center}
\end{figure}

\begin{figure}
\begin{center}
\includegraphics[width=.8\linewidth]{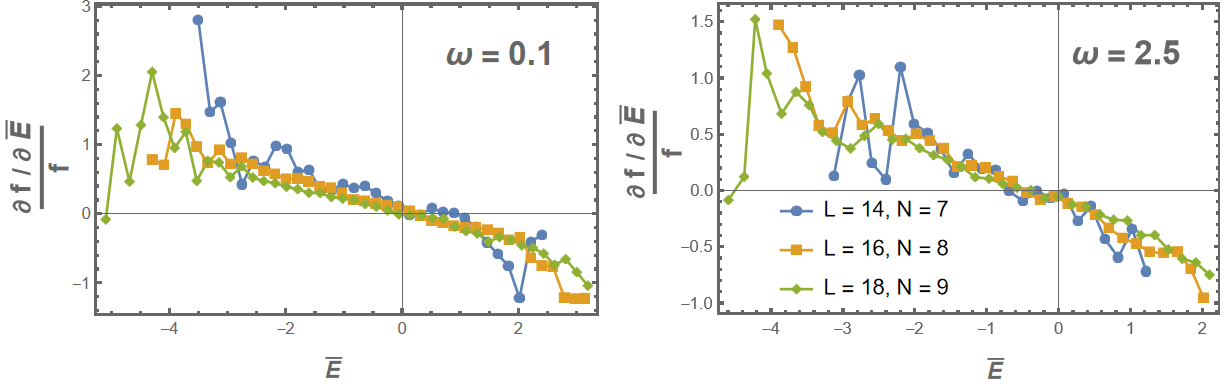}
\caption{\small{$1/f\partial f/\partial \bar{E}$ plots with fixed $\omega$'s for the spin current operator with different system sizes $L=14, N=7$, $L=16, N=8$, and $L=18, N=9$.}}
\label{dfdEb_Eb_innNNN_L14L16L18_omega0125_plot}
\end{center}
\end{figure}

Next we will study the normalized derivative $f'(\bar{E})/f(\bar{E})$. Figure \ref{dfdEb_Eb_innNNN_L14L16L18_omega0125_plot} are plots of $f'(\bar{E})/f(\bar{E})$ as a function of $\bar{E}$ for fixed values of $\omega$ for different system sizes. The slopes of the plots decrease with increasing system size. It is clearer now that $f'(\bar{E})/f(\bar{E})$ for fixed $\bar{E}$ and $\omega$ decreases as the system size increases.

\begin{figure}
\begin{center}
\includegraphics[width=.8\linewidth]{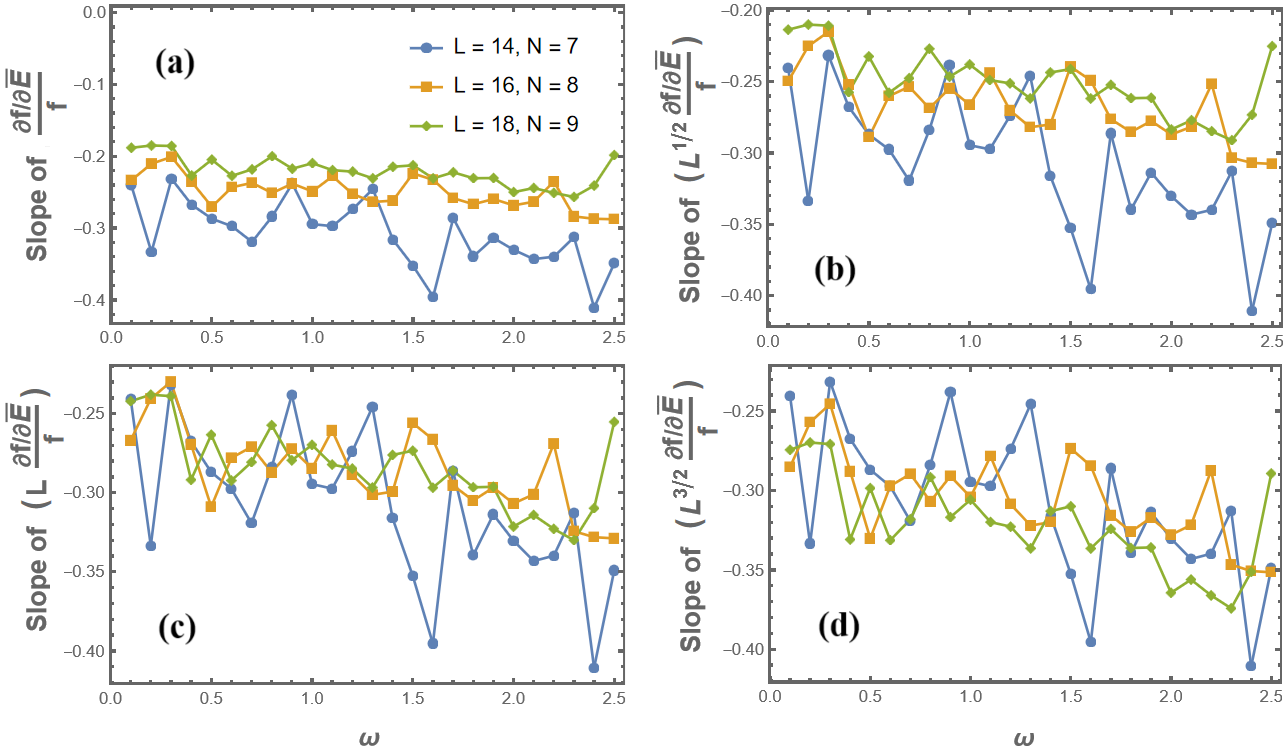}
\caption{\small{Panel (a) is plots of slope of linear fit of $1/f\partial f/\partial \bar{E}$ for $|\bar{E}|\leq 1$ for the spin current operator with different system sizes $L=14, N=7$, $L=16, N=8$, and $L=18, N=9$. Panels (b), (c) and (d) are plots of the slopes after scaling with different powers of the system size. Panel (c) shows that the slope is decreasing as $L^{-1}$ where $L$ is the system size.}}
\label{slopes_dfdEb_innNNN_plot}
\end{center}
\end{figure}

We can identify the scaling of the slope with system size $L$ (for a fixed $L/N$ ratio). For this, we plot the slopes of the linear fits of the $f'(\bar{E})/f(\bar{E})$ versus $\bar{E}$ plots for varying $\omega \in [0.1,2.5]$. To calculate the slopes reliably we use the $f'(\bar{E})/f(\bar{E})$ versus $\bar{E}$ plots with $\bar{E}\in [-1,1]$. Panel (a) in Figure \ref{slopes_dfdEb_innNNN_plot} is plots of the slope as a function of $\omega$. We find that the absolute value of the slope increases slightly with increasing $\omega$. What is more obvious is the flattening of $f$-function with increasing system size. Almost for the entire range of $\omega$, the slope of the largest system size $L=18, N=9$ is the lowest in absolute value. Panel (b), (c) and (d) are plots of the slope of the linear fits as a function $\omega$ with different system size scaling. Panel (c) clearly shows that the slopes of the different system sizes $L$ but scaled with $L^{1}$ mostly overlap. This implies that the $f$-function flattens as $L^{1}$. As mathematical expressions,
\begin{gather}
\frac{f'(\bar{E})}{f(\bar{E})}\;\xrightarrow{L\to\infty} \;0, \qquad L\;\frac{f'(\bar{E})}{f(\bar{E})}\; = \; \text{a constant.}
\label{ETHmono1}
\end{gather}
Numerical results for the kinetic energy operator and the single site spin operator in Appendix \ref{numres_app} further support these results. It is worthwhile to note that $\bar{E}$ is an extensive quantity. But the flattening of $f(\bar{E})$ is not merely a stretching of the function as we can see in Figure \ref{fEb_Eb_innNNNL18_omega0125_plot}.

\begin{figure}
\begin{center}
\includegraphics[width=.8\linewidth]{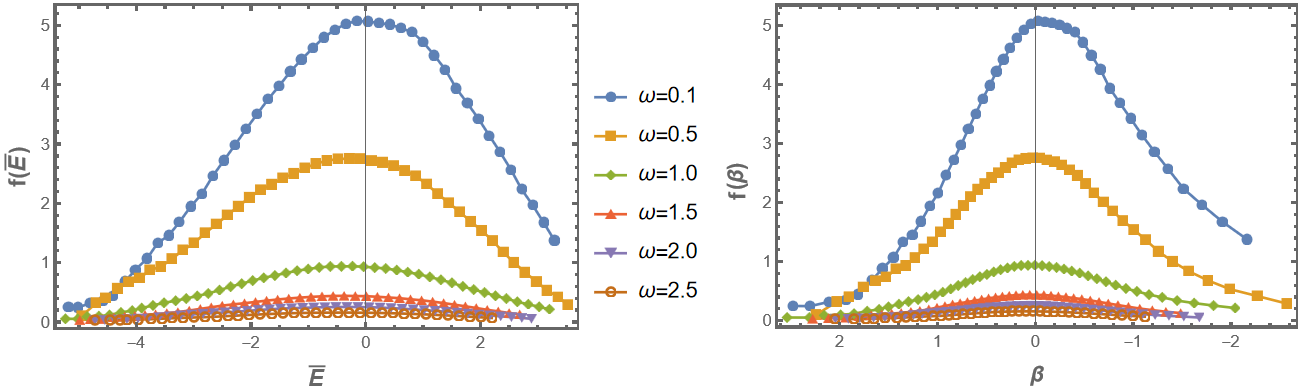}
\caption{\small{Left panel: $f(\bar{E})$ plots with different fixed $\omega$ of the spin current operator with system size $L=18$, $N=9$. Right panel: $f(\beta)$ plots with different fixed $\omega$ as a function of inverse effective temperature $\beta$ of the spin current operator with system size $L=18$, $N=9$.}}
\label{fEb_Ebbeta_innNNNL18_plot}
\end{center}
\end{figure}

In Figure \ref{fEb_Eb_innNNNL18_omega0125_plot}, we also observe an interesting property of the $f$-function. The variation of the $f$-function is not an arbitrary variation. We study this variation in detail for a fixed system size. For this, we work with the largest system size under consideration $L=18, N=9$. Figure \ref{fEb_Ebbeta_innNNNL18_plot} are plots of $f$-function as a function of $\bar{E}$ (left panel) and as function of the inverse temperature $\beta$ for different fixed values of $\omega$. We can see that $f$-function is a monotonically increasing function of the magnitude of the temperature $|T|$.

This monotonically increasing behaviour of $f$-function as a function of $|T|$ actually further reinforces the Kelvin statement, meaning the contribution from this variation is such that it gives extra support in making the change in energy $\Delta E$ to have the same sign as the temperature $T$ or the inverse temperature $\beta$.

Let us consider perturbations with $\omega$ cut-off at a finite value $\Lambda$. One example with a sharp cut-off \footnote{It may appear that this perturbation has infinite support from $t\to-\infty$ to $t\to\infty$ but one can introduce higher frequency modes with amplitudes much smaller than $\lambda_0$ and make $\lambda(t)\to 0$ faster than $1/t$. Nothing in the physical world has a sharp transition so it is always a competition of (energy) scales.} is
\begin{gather}
\lambda(t)=2\lambda_0\Lambda\,\text{sinc} (\Lambda t)\\
\tilde{\lambda}(\omega)=\lambda_0\,\text{rect} \left(\frac{\omega}{2\Lambda}\right)=\lambda_0\begin{cases} 0, &\quad \text{if} \quad |\omega|>\Lambda\\ \frac{1}{2}, &\quad \text{if} \quad |\omega|=\Lambda\\ 1, &\quad \text{if} \quad |\omega|<\Lambda\\ \end{cases}
\end{gather}
Considering a large enough system size $L$, we can make $\Lambda$ as small as we want without taking it smaller that the random matrix theory (RMT) scale $\sim 1/L^2$. We can also ensure that $\Lambda<|T|=1/|\beta|$. Then we can use the expansion of the exponential entropic factor in (\ref{microDE}). For small $\Lambda$, we consider only the leading order term which gives us
\begin{equation}
\Delta E_n=\int_{0}^{\Lambda}d\omega\, |\lambda(\omega)|^2\,\omega \left(f(\bar{E}+\omega/2,\omega)^2-f(\bar{E}-\omega/2,\omega)^2\right)\ 
\label{microDEsmall}
\end{equation}
where we have used $|\tilde{\lambda}(\omega)|=|\tilde{\lambda}(-\omega)|$ and $f(\bar{E},\omega)=f(\bar{E},-\omega)$. Now this quantity can be non-negative only if $f(\bar{E})$ is a monotonically increasing function of $\bar{E}$, which is the expected result for positive temperature. Similarly, we need $f(\bar{E})$ to be a monotonically decreasing function of $\bar{E}$ for negative temperature so that the above expression of $\Delta E_n$ is negative.

We want to emphasize that we are working in the small $\omega$ region so this argument is not valid for large $\omega$ region. More crucially, there is no reason why the above expression should be either positive or negative, unlike in subsection \ref{1stpass} where we invoked Berry's conjecture. But it is still a highly desirable result otherwise the chaotic system will violate the second law of thermodynamics in this low energy physics. Note that this leading $\omega^1$ term is absent in the thermodynamic limit because the $f$-function goes to the flat-limit of this monotonic behaviour.

Another motivation for the monotonically increasing behaviour of $f$-function as a function $|T|$ comes from the large $\omega$ behaviour of $f(\beta,\omega)$. As a function of $\omega$, it falls exponentially faster than $e^{-|\beta\omega|/4}$. This is indeed a monotonically increasing function of $|T|=1/|\beta|$.\footnote{We thank Anatoly Dymarsky for pointing out this result interpreted in terms of the monotonic behaviour of $f(\beta)$.} Again we want to emphasize that this exponential fall-off takes into account only $\omega$-dependence in $f(\beta,\omega)$. There could be other $\beta$-dependence even in the large $\omega$ region. But if we take large enough $\omega$, the other $\beta$-dependence cannot trump the exponential monotonic behaviour. So, at least we have this \emph{proof} of the monotonic behaviour for large enough $\omega$.

The entropy $S$ is a monotonically increasing function of the magnitude of the temperature $|T|$ in all reasonable physical systems. So, we have
\begin{center}
\emph{$f$-function is a monotonically increasing function of the magnitude of the temperature $|T(\bar{E})|$ or the entropy $S(\bar{E})$.}
\end{center}
This result along with (\ref{ETHmono1}) constitute ETH-monotonicity. We propose that every quantum chaotic many-body system obeying equivalence of ensembles will have this property.

\begin{figure}
\begin{center}
\includegraphics[width=.8\linewidth]{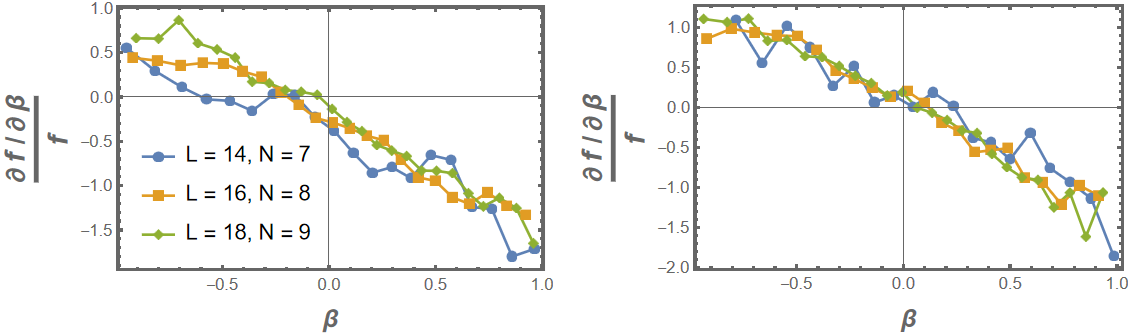}
\caption{\small{$1/f\partial f/\partial \beta$ plots with fixed $\omega$'s for the spin current operator with different system sizes $L=14, N=7$, $L=16, N=8$, and $L=18, N=9$. Unlike $f(\bar{E})$}, $f(\beta)$ does not flatten in the thermodynamic limit.}
\label{dfdbeta_beta_innNNN_L14L16L18_omega0125_plot}
\end{center}
\end{figure}

Note that temperature is an intensive quantity so the monotonically increasing behaviour will remain intact even in the thermodynamic limit. It can be verified from Figure \ref{dfdbeta_beta_innNNN_L14L16L18_omega0125_plot} which are plots of $1/f \partial f/\partial\beta$ as a function of $\beta$ for fixed values of $\omega$. Unlike $f(\bar{E})$, $f(\beta)$ does not flatten as the system size increases. But it is not physically relevant because what appeared in (\ref{microDE}) is $f$-function as a function of $\bar{E}$, not as a function of $T$ or $\beta$.

\section{Higher order terms of the change in energy}
\label{higher_sec}
We consider the higher order terms of the perturbation theory in this section. The $O(\lambda^3)$ term of the change in energy $\Delta E$ is
\begin{eqnarray}
\frac{i}{2}\sum_{l,m}E_l\left[\tilde{\lambda}(\omega_{nm})\tilde{\lambda}(\omega_{ml})\tilde{\lambda}(\omega_{ln})\mathcal{O}_{nm}\mathcal{O}_{ml}\mathcal{O}_{ln}-\tilde{\lambda}(\omega_{nl})\tilde{\lambda}(\omega_{lm})\tilde{\lambda}(\omega_{mn})\mathcal{O}_{nl}\mathcal{O}_{lm}\mathcal{O}_{mn}\right]\
\end{eqnarray}
where $\omega_{nm}=E_n-E_m$. This term is identically zero because $\tilde{\lambda}(\omega)=\tilde{\lambda}(-\omega)^*$, $\mathcal{O}_{nm}=\mathcal{O}^*_{mn}$ and the sums run over the entire state space. Similarly, all odd power terms of  $\lambda(t)$ are identically zero.

The $O(\lambda^4)$ term of the change in energy $\Delta E$ is
\begin{equation}
\frac{1}{4}\,\sum_{k,m,l} (E_n+E_l-E_k-E_m)[\tilde{\lambda}(\omega_{nk})\tilde{\lambda}(\omega_{kl})\tilde{\lambda}(\omega_{lm})\tilde{\lambda}(\omega_{mn})\mathcal{O}_{nk}\mathcal{O}_{kl}\mathcal{O}_{lm}\mathcal{O}_{mn}]\
\end{equation}
In general, we find that this sum is negative (positive, zero) for initial states with positive (negative, infinite) temperature. $R_{mn}$ are not completely random \cite{Foini:2018sdb}. If they were completely random, these higher order terms would be suppressed by exponentials of the entropy and a full analytic proof of the arrow of time at all orders of $\lambda(t)$ would be possible.

\begin{figure}[h]
\begin{center}
\includegraphics[width=.8\columnwidth]{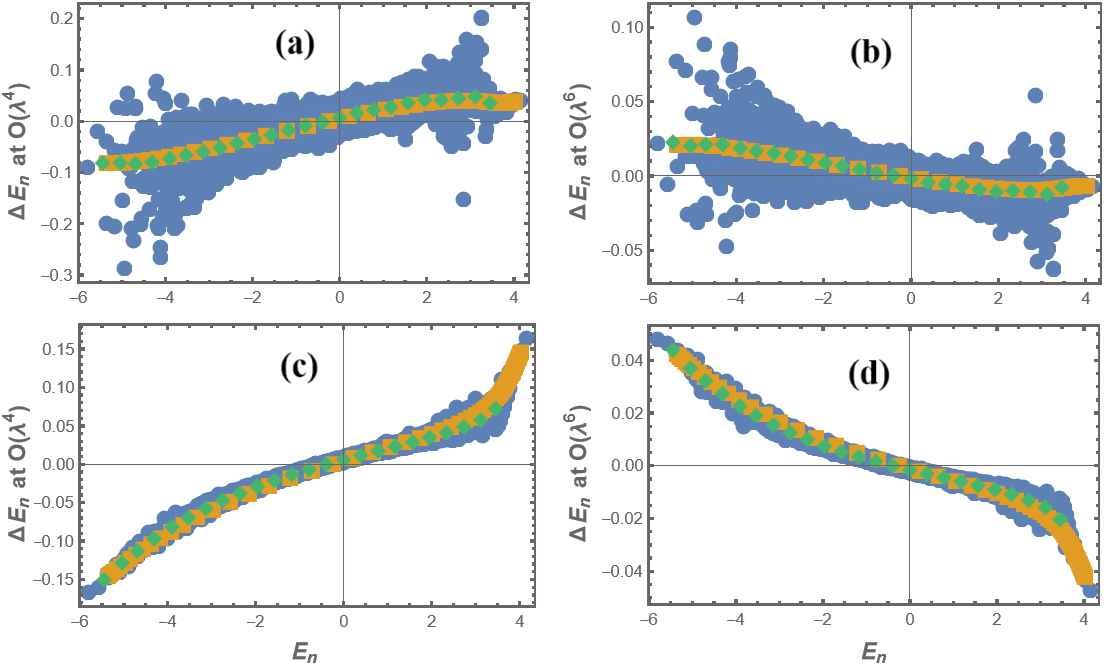}
\caption{\small{$\lambda(t)^4$ (left panels) and $\lambda(t)^6$ (right panels) order in $\Delta E$ series expansion for the chaotic spin chain ($L=16, N=8$) after a unit delta function perturbation $\lambda(t)=\delta(t)$. The blue dots are for the different energy eigenstates of energy $E_n$ as the initial state, the yellow plots are for canonical ensembles and the green dots are for microcanonical ensembles. The top panels are for perturbation with the spin current operator and the bottom panels are for perturbations with the kinetic energy operator.}}
\label{l4-l6_XXZ}
\end{center}
\end{figure}

In general, the even powers of $\lambda(t)$ have definite signs (statistically speaking) and the higher order terms are suppressed. Figure \ref{l4-l6_XXZ} are plots of $\lambda(t)^4$ and $\lambda(t)^6$ terms in $\Delta E$ series expansion for the chaotic XXZ spin chain ($L=16, N=7$) after a unit delta function perturbation $\lambda(t)=\delta(t)$. We can see that these higher order terms are smaller (roughly by a factor of $2$ for the quartic term) compared to the leading order term even with a unit perturbation strength.

\section{Large perturbations}
\label{lperturb}
\begin{figure}
\begin{center}
\includegraphics[width=.8\columnwidth]{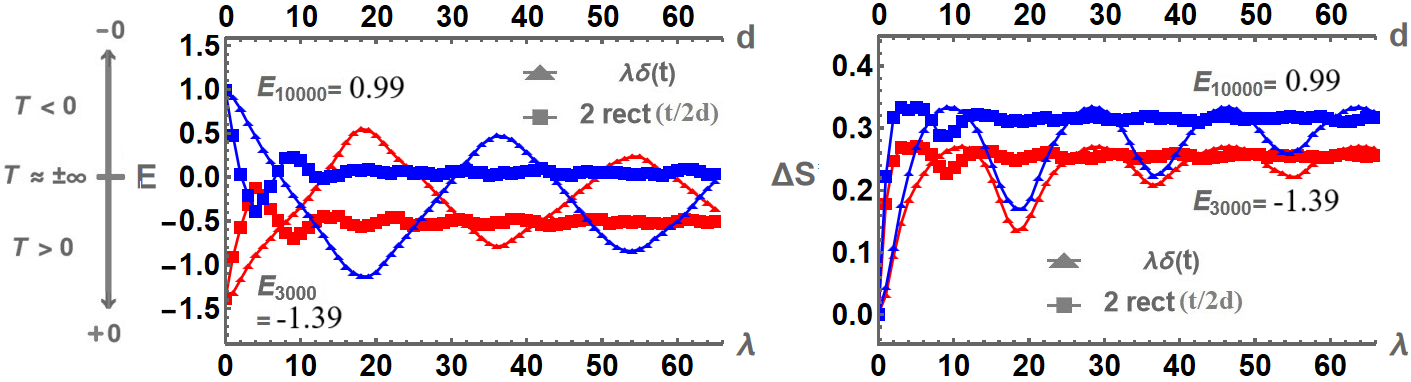}
\caption{\small{Left: Change in energy after large perturbations in the chaotic spin chain of size $L=16, N=8$. Starting from the positive temperature state, the system always gains energy (red curves). Starting from the negative temperature state, the system always loses energy (blue curves). Right: The change in canonical entropy is always positive.}}
\label{large_perturbations}
\end{center}
\end{figure}

We also study perturbations with large values of $\lambda(t)$ numerically. We considered perturbations of the form $\lambda(t)=\lambda \delta(t)$ and $\lambda(t)=2 \;\text{rect} (t/2d)$ which is the rectangular function of height $2$ and width $2d$. $\Delta E$ and $\Delta S$ as a function of $\lambda$ and $d$ are plotted in Figure \ref{large_perturbations}. We have considered the initial eigenstates $|E_{3000}\rangle$ with positive temperature and $|E_{10000}\rangle$ with negative temperature. $\Delta E$ remains always positive (negative) for positive (negative) initial temperature. This agrees with the Kelvin form of the second law of thermodynamics. The change in canonical entropy is also always positive in these examples. Note that the energy even jumps across the infinite temperature region.

We have not studied the time evolution. We calculate the change in entropy by matching the final energy with the canonical expectation value of the Hamiltonian. With large perturbations, we cannot use the formula $\Delta E=T\Delta S$.

\section{Conclusions}
\label{concl}
We show that quantum chaotic many-body systems inherently possess an arrow of time in the form of the Kelvin statement of the second law of thermodynamics. Working at leading order in perturbation theory, we show that the arrow of time also constrains the off-diagonal terms in the ETH statement. The new constraints have to be taken into account when one is working beyond probe limit, that is, when the effect or feedback of the probe on a quantum system cannot be ignored. The constraints are on the $f$-function as a function of $\bar{E}$ in the ETH statement.
\begin{enumerate}
\item $f$-function is a constant function of $\bar{E}$ in the thermodynamic limit.
\item $f$-function is a monotonically increasing function of the magnitude of the temperature $|T(\bar{E})|$ or the entropy $S(\bar{E})$.
\item $f(\bar{E})$ flattens as $L^{1}$ as the system size $L$ increases.
\end{enumerate}
We call these properties of $f(\bar{E})$ collectively as ETH-monotonicity. These constraints are also constraints on the transition amplitudes between energy eigenstates.

We also calculated the higher order terms of $\lambda(t)$ in $\Delta E$ numerically. For reasonable perturbation strength, the higher order terms are small compared to the leading term. The odd power terms of $\lambda(t)$ are identically zero.

We can still reduce the entropy by using a fine-tuned perturbation. This can be accomplished by predominantly turning on modes $\tilde{\lambda}(\omega)$ which allows transition to only lower energy levels. But this fine-tuning requires the knowledge of a large part of the spectrum and precise knowledge of the initial state. This is reminiscent of the Maxwell's demon problem in classical thermodynamics. In the present case, finding the energy levels at the required precision involves performing highly precise measurements which increases the total entropy of the system and the outside agent.

In a finite system, for which $f(\bar{E})$ is not flat, the monotonic behaviour of $f(\bar{E})$ enhances or reinforces the arrow of time as we have seen in (\ref{microDEsmall}). For example, from the $\omega^1$ term of that equation, the system  at positive temperature appears to absorb more energy per system size when the system size is small compared to a larger system, for same strength of the perturbation. It would be interesting to see if this result can be considered as a quantum advantage. We wish to study further in this direction in future.

It would also be interesting to check for ETH-monotonicity in long-range interacting systems for which equivalence of ensembles breaks down \cite{Barr__2001}.

\section*{Acknowledgements}
The author would like to thank Gautam Mandal for helpful discussions during one of which the main idea for this paper arose.

\paragraph*{Funding information}
The author is fully supported by the Department of Science and Technology (Government of India) under the INSPIRE Faculty fellowship scheme (IFA-20-PH-262).

\appendix

\section{Change in energy starting from an energy eigenstate}
\label{dEn_app}
The change in energy (\ref{dE1}) is written in Schrodinger picture. In interaction picture, it can be written as
\begin{equation}
\Delta E_n=\langle n|\bar{U}_I(t_f)H_0 U_I(t_f)|n\rangle-\langle n|H_0|n\rangle\
\label{dE2}
\end{equation}
where $\bar{U}_I(t_f)$ and $U_I(t_f)$ are the time evolution operators in interaction picture. Note that this simplified expression is possible because we are calculating the expectation value of $H_0$ and $e^{it H_0}H_0e^{-it H_0}=H_0$. The Dyson series expansion of the time evolution operators are
\begin{eqnarray}
\bar{U}_I(t_f)&=&1+i\int_{t_i}^{t_f}dt_1\,\lambda(t_1)\mathcal{O}_I(t_1)-\int_{t_i}^{t_f}dt_1\int_{t_i}^{t_1}dt_2\,\lambda(t_1)\lambda(t_2)\mathcal{O}_I(t_2)\mathcal{O}_I(t_1)\nonumber\\
&&-i\int_{t_i}^{t_f}dt_1\int_{t_i}^{t_1}dt_2\int_{t_i}^{t_2}dt_3\,\lambda(t_1)\lambda(t_2)\lambda(t_3)\mathcal{O}_I(t_3)\mathcal{O}_I(t_2)\mathcal{O}_I(t_1)+O(\lambda^4)\nonumber\\
U_I(t_f)&=&1-i\int_{t_i}^{t_f}dt_1\,\lambda(t_1)\mathcal{O}_I(t_1)-\int_{t_i}^{t_f}dt_1\int_{t_i}^{t_1}dt_2\,\lambda(t_1)\lambda(t_2)\mathcal{O}_I(t_1)\mathcal{O}_I(t_2)\nonumber\\
&&+i\int_{t_i}^{t_f}dt_1\int_{t_i}^{t_1}dt_2\int_{t_i}^{t_2}dt_3\,\lambda(t_1)\lambda(t_2)\lambda(t_3)\mathcal{O}_I(t_1)\mathcal{O}_I(t_2)\mathcal{O}_I(t_3)+O(\lambda^4)\nonumber\\
&&\hspace{2.5cm}\mathcal{O}_I(t)=e^{i(t-t_i)H_0}\mathcal{O}(t_i)e^{-i(t-t_i)H_0}\nonumber\
\end{eqnarray}
$\bar{U}$ expansion is anti-time ordered and $U$ expansion is time-ordered. $\mathcal{O}_I(t)$ is the perturbing operator in interaction picture. With these expressions, the non-zero leading order term in (\ref{dE2}) is
\begin{eqnarray}
\Delta E_n&=&\int_{t_i}^{t_f}dt_1dt'_1\,\lambda(t_1)\lambda(t'_1)\,\langle n|\mathcal{O}_I(t_1)H_0\mathcal{O}_I(t'_1)|n\rangle\nonumber\\
&&-\int_{t_i}^{t_f}dt_1\int_{t_i}^{t_1}dt_2\,\lambda(t_1)\lambda(t_2)\,\langle n|\mathcal{O}_I(t_2)\mathcal{O}_I(t_1)H_0+H_0\mathcal{O}_I(t_1)\mathcal{O}_I(t_2)|n\rangle\
\end{eqnarray}
For brevity, we use a single integral sign for same type of integrals with equal limits of integration. Without any lost of generality, we can take $t_i\to-\infty$ and $t_f\to+\infty$ because we are considering $\lambda(t)$ with a finite support.
We used the Fourier transform relations
\begin{equation}
\tilde{\lambda}(\omega)=\int_{-\infty}^{\infty} dt\,\lambda(t) e^{it\omega}, \qquad \lambda(t)=\int_{-\infty}^{\infty} \frac{d\omega}{2\pi}\,\tilde{\lambda}(\omega) e^{-it\omega}\
\label{Fourierdef}
\end{equation}
Inserting a complete set of projection operators $\sum_m|E_m\rangle\langle E_m|$ and evaluating the expectation values we get
\begin{gather*}
\sum_m |\mathcal{O}_{nm}|^2\left[E_m\int_{-\infty}^{\infty}dt_1dt'_1\int_{-\infty}^{\infty}\frac{d\omega_1 d\omega'_1}{(2\pi)^2}\,\tilde{\lambda}(\omega_1)\tilde{\lambda}(\omega'_1)\,e^{it_1(E_n-E_m-\omega_1)}e^{it'_1(E_m-E_n-\omega'_1)}\right.\\
\left. -E_n\int_{-\infty}^{\infty}dt_1\int_{-\infty}^{t_1}dt_2\int_{-\infty}^{\infty}\frac{d\omega_1 d\omega_2}{(2\pi)^2}\,\tilde{\lambda}(\omega_1)\tilde{\lambda}(\omega_2)\left\{e^{it_2(E_n-E_m-\omega_2)}e^{it_1(E_m-E_n-\omega_1)}+(t_1 \leftrightarrow t_2)\right\}\right]\
\end{gather*}
The term inside the square bracket in the first line can be easily evaluated. The second line requires a few more steps. $\lambda(t)$'s are replaced by their Fourier transforms. Evaluating the $t_2$ integrals with $i\epsilon$ prescription and evaluating the $t_1$ integrals, the second line becomes
\begin{eqnarray}
&&-E_n\int_{-\infty}^{\infty}\frac{d\omega_1 d\omega_2}{2\pi}\,\tilde{\lambda}(\omega_1)\tilde{\lambda}(\omega_2)\left\{\frac{\delta(\omega_1+\omega_2)}{i(E_n-E_m-\omega_2-i\epsilon)}+\frac{\delta(\omega_1+\omega_2)}{i(E_m-E_n-\omega_2-i\epsilon)}\right\}\nonumber\\
&=&-E_n|\tilde{\lambda}(E_m-E_n)|^2\
\label{iep_int}
\end{eqnarray}
Each term inside the parenthesis in the first line contributes half of the final result. The $\omega_1$ integrals are trivial. We remind here again that $\tilde{\lambda}(\omega)=\tilde{\lambda}(-\omega)^*$ because $\lambda(t)$ is a real function. The $\omega_2$ integrals are evaluated using Sokhotski–Plemelj theorem over the real line. The two Cauchy principal value integrals cancelled out because $|\tilde{\lambda}(\omega)|^2$ is an even function. Substituting these results, we get (\ref{dEl2}).

\section{Change in energy starting from a thermal state}
\label{dEtherm_app}
We can repeat the calculation of \ref{dEn_app} for the case when the initial state is a thermal state defined by a density matrix $\rho_0=e^{-\beta H_0}$. The change in energy is given by
\begin{equation}
\Delta E_{\beta}=\frac{1}{Z}\text{Tr} \left(U_I(t_f)\rho_0 \bar{U}_I(t_f)\,H_0\right)-\frac{1}{Z}\text{Tr} \left(\rho_0 H_0\right), \qquad Z=\text{Tr} \rho_0\
\end{equation}
But to make it clear that the signs of the change in energy is as expected from second law, we will perform the calculation in a slightly different manner. Again note that the above simplified expression in the interaction picture is possible because we are calculating the expectation value of $H_0$ and $e^{it H_0}H_0e^{-it H_0}=H_0$. Using the Dyson series expansions, we get
\begin{eqnarray}
\Delta E_{\beta}&=&\frac{1}{Z}\int_{t_i}^{t_f}dt_1dt'_1\,\lambda(t_1)\lambda(t'_1)\;\sum_n \;\langle n|\mathcal{O}_I(t_1)e^{-\beta H_0}\mathcal{O}_I(t'_1)H_0|n\rangle\nonumber\\
&&-\frac{1}{Z}\int_{t_i}^{t_f}dt_1\int_{t_i}^{t_1}dt_2\,\lambda(t_1)\lambda(t_2)\;\sum_n\;\langle n| e^{-\beta H_0}\left(\mathcal{O}_I(t_2)\mathcal{O}_I(t_1)H_0+H_0\mathcal{O}_I(t_2)\mathcal{O}_I(t_1)\right)|n\rangle\nonumber\
\end{eqnarray}
Note the ordering of the operators in the first line where we have used the cyclic property of trace. The expectation values are evaluated by inserting a complete set of projection operators $\sum_m|E_m\rangle\langle E_m|$. In the first line, we split the single term into two halves and interchanged the indices $n$ and $m$ for the second half. In the evaluation of the expectation values of the two operator products in the second line, we also interchange the indices $n$ and $m$ for the second operator product. These steps give us
\begin{eqnarray}
\Delta E_{\beta}&=&\frac{1}{Z}\int_{t_i}^{t_f}dt_1dt'_1\,\lambda(t_1)\lambda(t'_1)\left[\frac{1}{2}\,\sum_{n,m} E_n e^{-\beta E_m} |\mathcal{O}_{nm}|^2e^{it_1(E_n-E_m)}e^{it'_1(E_m-E_n)}\right.\nonumber\\
&&\hspace{5cm} +\left.\frac{1}{2}\,\sum_{n,m} E_m e^{-\beta E_n} |\mathcal{O}_{nm}|^2e^{it_1(E_m-E_n)}e^{it'_1(E_n-E_m)}\right]\nonumber\\
&&-\frac{1}{Z}\int_{t_i}^{t_f}dt_1\int_{t_i}^{t_1}dt_2\,\lambda(t_1)\lambda(t_2)\left[\sum_{n,m}E_n e^{-\beta E_n}|\mathcal{O}_{nm}|^2e^{it_2(E_n-E_m)}e^{it_1(E_m-E_n)}\right.\nonumber\\
&&\hspace{5cm} +\left.\sum_{n,m}E_m e^{-\beta E_m}|\mathcal{O}_{nm}|^2e^{it_1(E_m-E_n)}e^{it_2(E_n-E_m)}\right]\
\label{dEtherm_eqn}
\end{eqnarray}
Now consider the first and the second lines. $\lambda(t)$'s are replaced by their Fourier transforms. Taking the limits $t_i\to-\infty$ and $t_f\to+\infty$, and performing the $t$ integrals give us
\begin{eqnarray}
&&\frac{1}{2Z}\int_{-\infty}^{\infty}d\omega_1 d\omega'_1\,\tilde{\lambda}(\omega_1)\tilde{\lambda}(\omega'_1)\sum_{n,m} |\mathcal{O}_{nm}|^2 \left[E_n e^{-\beta E_m} \delta(\omega_1-(E_n-E_m))\delta(\omega'_1-(E_m-E_n))\right.\nonumber\\
&&\hspace{5cm}\left. +E_m e^{-\beta E_n} \delta(\omega_1-(E_n-E_m))\delta(\omega'_1-(E_m-E_n))\right]\nonumber\\
&=&\frac{1}{2Z}\int_{-\infty}^{\infty}d\omega\,|\tilde{\lambda}(\omega)|^2\sum_{n,m} \left(E_n e^{-\beta E_m}+E_m e^{-\beta E_n}\right) |\mathcal{O}_{nm}|^2\delta(\omega-(E_m-E_n))\
\end{eqnarray}

The integrals in the third and fourth lines in (\ref{dEtherm_eqn}) are evaluated as in section \ref{dEn_app}. The Cauchy principal values again cancelled out because $n$ and $m$ run over the entire state space, and $|\tilde{\lambda}(\omega)|^2$ is an even function. So, the third and fourth lines in (\ref{dEtherm_eqn}) become
\begin{eqnarray}
-\frac{1}{2Z}\int_{-\infty}^{\infty}d\omega\,|\tilde{\lambda}(\omega)|^2\sum_{n,m} \left(E_n e^{-\beta E_n}+E_m e^{-\beta E_m}\right) |\mathcal{O}_{nm}|^2\delta(\omega-(E_m-E_n))\
\end{eqnarray}
Finally, the change in energy is given by
\begin{eqnarray}
\Delta E_{\beta}&=&\frac{1}{2Z}\int_{-\infty}^{\infty}d\omega\,|\tilde{\lambda}(\omega)|^2\sum_{n,m} \left(E_m-E_n\right)\left(e^{-\beta E_n}-e^{-\beta E_m}\right) |\mathcal{O}_{nm}|^2\delta(\omega-(E_m-E_n))\nonumber\\
&=&\frac{1}{2}\int_{-\infty}^{\infty}d\omega\,\omega |\tilde{\lambda}(\omega)|^2 \left[\frac{\left(e^{\beta \omega}-1\right)}{Z}\sum_{n,m}  e^{\beta E_m} |\mathcal{O}_{nm}|^2\delta(\omega-(E_m-E_n))\right]\nonumber\
\label{DEbetagen}
\end{eqnarray}
This quantity is always a positive (negative or zero) at positive (negative or infinite) temperature $1/\beta$. The quantity inside the square brackets as a function of $\omega$ is the spectral function $A(\omega)$ \cite{Zubarev_1960,sglrbook2013,coleman15} in a thermal state.\footnote{The quantity inside the square brackets in (\ref{DEbetagen}) without the $\left(e^{\beta \omega}-1\right)$ factor is called spectral intensity in \cite{Zubarev_1960}.}
\begin{eqnarray}
A(\omega)&=&\frac{1}{Z}\left(e^{\beta \omega}-1\right)\sum_{n,m}  e^{-\beta E_m} |\mathcal{O}_{nm}|^2\delta(\omega-(E_m-E_n))\nonumber\\
&=&-\frac{1}{Z}\left(e^{-\beta \omega}-1\right)\sum_{n,m}  e^{-\beta E_n} |\mathcal{O}_{nm}|^2\delta(\omega-(E_m-E_n))\
\label{spectralfn}
\end{eqnarray}
It is an odd function of its argument $\omega$. It is also the imaginary part of the retarded Green function, which is a relation that will become useful when we rederive the change in energy below using linear response theory.

\subsection{Derivation using linear response theory}
\label{dEthermLRT_app}
To make the connection with linear response theory (LRT) clear, we rederive the change in energy using Kubo formula. We consider the time dependent Hamiltonian (\ref{Ht}) where the real function $\lambda(t)$ has support for a finite range $-d<t<d$. The initial density matrix $\rho=e^{-\beta H_0}$ satisfies $[\rho,H_0]=0$. The rate of change of total energy \cite{jjmrbook1991} is given by
\begin{equation}
\frac{dE}{dt}=\frac{d}{dt}\langle \mathcal{H}\rangle=\left\langle \frac{d\mathcal{H}}{dt}\right\rangle=\dot{\lambda}(t)\langle\mathcal{O}(t)\rangle\
\label{dEdt}
\end{equation}
where $\dot{\lambda}(t)=d\lambda/dt$ and $\mathcal{O}(t)$ is the perturbing operator in Heisenberg picture. The $d\rho/dt$ term can be shown to be zero using the time evolution equation of $\rho$ and the cyclic property of the trace. We are working with LRT so we will expand $\mathcal{O}(t)$ only upto linear order in $\lambda(t)$.
\begin{gather}
\langle\mathcal{O}(t)\rangle=\langle\mathcal{O}(0)\rangle+\int_{t_i}^{\infty}dt'\,\lambda(t')G_R(t,t')\\
G_R(t,t')=-i\theta(t-t')\langle[\mathcal{O}(t),\mathcal{O}(t')]\rangle\
\end{gather}
$G_R$ is the definition of retarded Green function (in general cases, also called response function) of the operator $\mathcal{O}$.  This relation is also known as the Kubo formula for the response function \cite{Kubo:1957mj,jjmrbook1991}. Considering $t_i<-d$ and $d<t_f$, the total energy gained upto this order is
\begin{gather}
\Delta E_{\beta}=\int_{t_i}^{t_f}dt\,\frac{dE}{dt}=\langle\mathcal{O}(0)\rangle\int_{t_i}^{t_f} dt\,\dot{\lambda}(t) +\int_{t_i}^{t_f}dt\,\dot{\lambda}(t)\int_{t_i}^{\infty} dt' \, G_R(t,t')\lambda(t')\
\end{gather}
The first term is zero because $\lambda(t_i)=\lambda(t_f)=0$ and $\langle\mathcal{O}(0)\rangle$ is a constant - the initial thermal expectation value of the operator $\mathcal{O}$.
We can take $t_i\to-\infty$ and $t_f\to\infty$ because $\lambda(t)$ has only a finite support. $G_R(t,t')=G_R(t-t')$ because it is calculated in the initial thermal state. Now after Fourier transformations we get,
\begin{eqnarray}
\Delta E_{\beta}=i\int_{-\infty}^{\infty} d\omega \omega |\lambda(\omega)|^2\; \tilde{G}_R\left(\omega\right)=\frac{1}{2}\int_{-\infty}^{\infty} d\omega \omega |\tilde{\lambda}(\omega)|^2\; A\left(\omega\right)\
\end{eqnarray}
The real part of $\tilde{G}_R(\omega)$ is even and the imaginary part of $\tilde{G}_R(\omega)$ is odd and $A(\omega)=-2\,\text{Im}\;\tilde{G}_R(\omega)$.

\section{Comparison of canonical and microcanonical ensembles}
\label{cmccomp_app}
In this section, we will compare canonical and microcanonical ensembles for different system sizes $L = 14, N=7$, and $L = 16, N=8$, and $L = 18, N=9$ of the chaotic XXZ spin chain. $L$ is the total number of spin and $N$ is the number of up-spin.  The corresponding dimensions of the state space are $\binom{14}{7}=3432$, $\binom{16}{8}=12870$, and $\binom{18}{9}=48620$. We compare the thermal entropy calculated using different ensembles.

\begin{figure}
\begin{center}
\includegraphics[width=.8\linewidth]{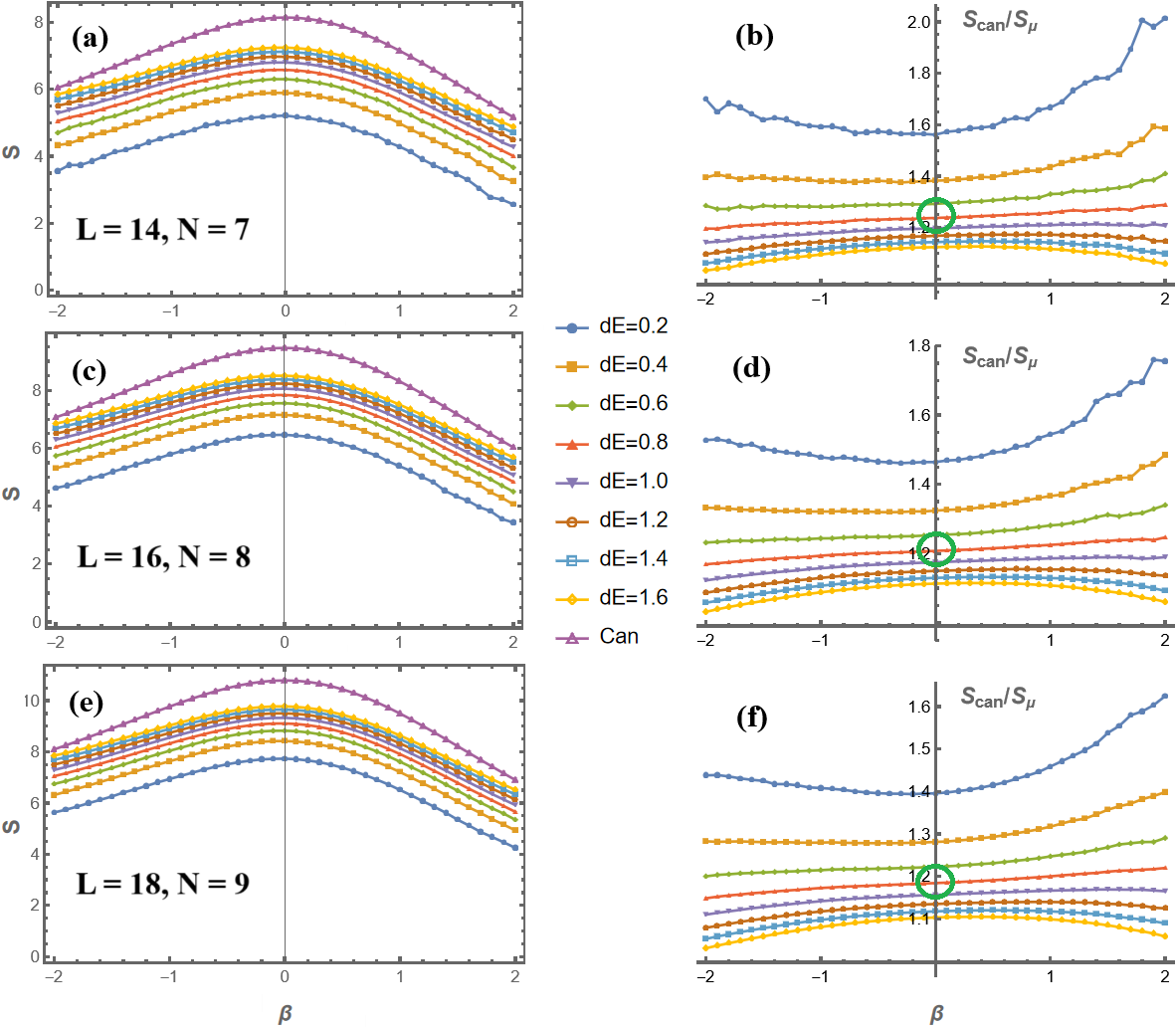}
\caption{\small{Left panels: Comparison of canonical entropy $S_{\text{can}}$ and microcanonical entropy $S_{\mu}$ as a function of inverse temperature $\beta$ for different system sizes of chaotic spin chain using different values of the energy window $dE$. Right panels: Plots of $S_{\text{can}}/S_{\mu}$ as a function of inverse temperature $\beta$ for different values of $dE$ to study $dE$-dependence of the microcanonical entropy and to find the optimum value of $dE \sim 0.8$.}}
\label{cmc_comp}
\end{center}
\end{figure}

In the canonical ensemble, we study inverse temperature range $\beta\in [2,-2]$. We calculate the canonical entropy using the formula
\begin{equation}
S_{\text{can}}=-\sum_{i} p_i\log p_i, \qquad p_i=e^{-\beta E_i}
\end{equation}
The sum is over the total dimension of the state space. For a particular value of $\beta$, we again calculate the entropy of microcanonical ensemble by counting the number of energy eigenstates in an energy window $dE$ centered at the thermal value of the energy $E_{\beta}=\sum_i p_iE_i$.
\begin{equation}
S_{\mu}=\log W, \qquad \text{$W=$ number of energy eigenstates within $[E_{\beta}-dE/2,E_{\beta}+dE/2]$}
\end{equation}
It is generally considered that $dE$ is arbitrary. A critical requirement is that $dE$ should be subextensive. Below we will identify an optimum value of $dE$. We calculate $S_{\mu}$ with different values of $dE$. The left panels of Figure \ref{cmc_comp} are the plots of $S_{\text{can}}$ and $S_{\mu}$ calculated using different values of $dE\in[0.2,1.6]$. We find that the entropy calculated from the canonical ensembles is always greater than the corresponding microcanonical entropy. With larger value of $dE$, the microcanonical entropy tends towards the canonical entropy but the difference in the $\beta$ dependence remains. 

\begin{figure}
\begin{center}
\includegraphics[width=.8\linewidth]{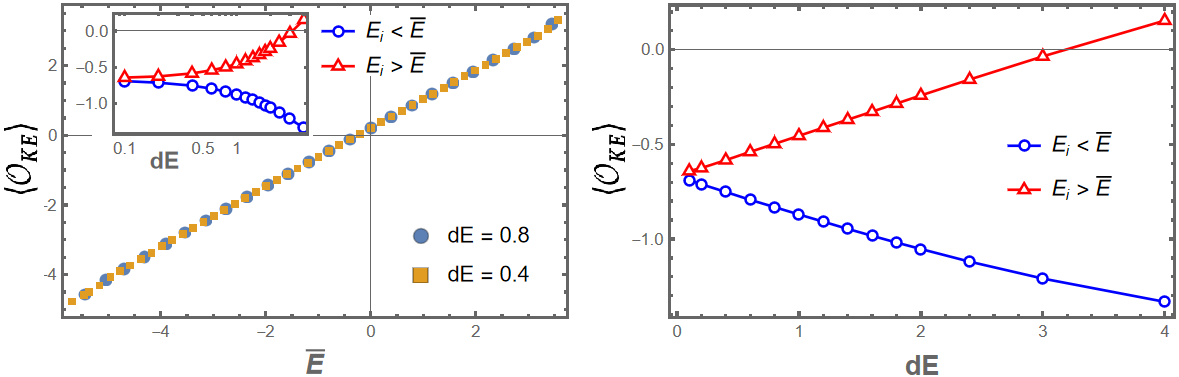}
\caption{\small{Finding the value of the energy window $dE$ following \cite{Rigol_2008nature}. These plots are for the system size $L=16, N=8$. Left panel: Plot of microcanonical expectation values of the kinetic energy operator using two different sizes of the energy window $dE=0.4$ and $dE=0.8$. Inset is the Log-Linear plot of the expectation value of the left ($[\bar{E}-dE/2,\bar{E}]$) and right $[\bar{E},\bar{E}+dE/2]$ averages as a function of $dE$, where $\bar{E}=-1.05$ which has effective temperature $T=3$. Right panel: Linear-Linear plot of the expectation value of the kinetic energy operator in the half energy window at left ($[\bar{E}-dE/2,\bar{E}]$) and at right $[\bar{E},\bar{E}+dE/2]$ as a function of $dE$, where $\bar{E}=-1.05$ which has effective temperature $T=3$.}}
\label{kedE04dE08}
\end{center}
\end{figure}

Since the difference between canonical entropy and microcanonical entropy will always be present for a finite system, the next thing we can attempt is whether we can ensure that the ratio between canonical entropy and microcanonical entropy remains constant with varying temperature or energy. While working with different system sizes, it is naturally better to use temperature instead of energy. 
So to identify an optimum choice of $dE$, we study the ratio $S_{\text{can}}/S_{\mu}$ as a function of inverse temperature $\beta$ using different values of $dE$. The right panels in Figure \ref{cmc_comp} are the plots of the ratio. We look for $\beta$-independence from the plots using different values of $dE$. For small $dE$, the ratio is a convex function of $\beta$. For large $dE$, the ratio is a concave function. So, there is indeed an optimum value of $dE$ for which the ratio is approximately independent of $\beta$. From the plots, we can see that the optimum value of $dE$ is $dE=0.8$. This optimum value of $dE$ is also independent of the system size.

For the optimum value of $dE$, the ratio $S_{\text{can}}/S_{\mu}$ is around $1.2$. So, for the system sizes under consideration, the canonical and microcanonical entropies are within 20\% of each others' values. The ratio is closer to one for the higher dimensional state space. We point this out using the green circles in Figure \ref{cmc_comp}. For example, for the system $L=14, N=7$, the ratio with the optimum value of $dE$ is close to $1.22$ at infinite temperature $\beta=0$. Whereas, the corresponding value is close to $1.18$ for the system $L=18, N=9$. This is expected because we will see the equivalence of ensembles $S_{\text{can}}/S_{\mu}\to 1$ in the thermodynamic limit.

The choice of optimum $dE$ arises from general consideration. It is also motivated by our numerical calculation of $f(\bar{E},\omega)$. We have used the microcanonical entropy in the calculations of $f(\bar{E},\omega)$. Now with the choice of optimum $dE$, even if we had used the canonical entropy, our numerical results supporting ETH-monotonicity remain robust. If we had used canonical entropy instead of microcanonical entropy, the plots of $f(\bar{E},\omega)$ will be rescaled by an overall factor of $\sim e^{.1 S_{\mu}(\bar{E})}$. Now this factor in strictly a monotonically increasing function of absolute value of the temperature $|T|$. So, it will only reinforce the monotonically increasing nature of $f$-function as a function of the magnitude of the temperature $|T|$ or the entropy $S$.

Now we compare our choice of the energy window with the energy window used in \cite{Rigol_2008nature}. We consider the system size $L=16, N=8$. In the left panel of Figure \ref{kedE04dE08}, we plotted the microcanonical expectation values of the kinetic energy operator using two different sizes of the energy window. It appears that $dE=0.4$ and $dE=0.8$ are equally good. The inset of the same figure is the Log-Linear plot (following \cite{Rigol_2008nature}) of the expectation value in the half energy window at left ($[\bar{E}-dE,\bar{E}]$)and at right $[\bar{E},\bar{E}+dE]$ as a function of $dE$, where $\bar{E}=-1.05$ which has temperature $T=3$. It appears that our choice of $dE$ is a reasonable choice from this plot. But this depends on the choice of the range of $dE$ used. In other word, we believe that choosing a $dE$ from this Log-Linear plot is arbitrary. If we plot a Linear-Linear plot instead of a Log-Linear plot, then we get the right panel of Figure \ref{kedE04dE08}. Now, it appears that every choice of $dE$ is equally good (or bad) as long as $dE$ is not very large and comparable to the total size of the energy spectrum.

In conclusion, we believe that our choice of $dE$ is more natural and it has a precise definition. As we have seen above, it also does not scale extensively with the system size. It is rather a constant for the system sizes under consideration.
Note that we only need the comparison of canonical entropy and microcanonical entropy for a single fixed system size to find $dE$. We are comparing the optimum $dE$ for different system sizes because we anyway have to work with different system sizes for other results in the main text.

\section{The entropic factor in finite dimensional state space}
\label{entfac_app}
\begin{figure}
\begin{center}
\includegraphics[width=.6\linewidth]{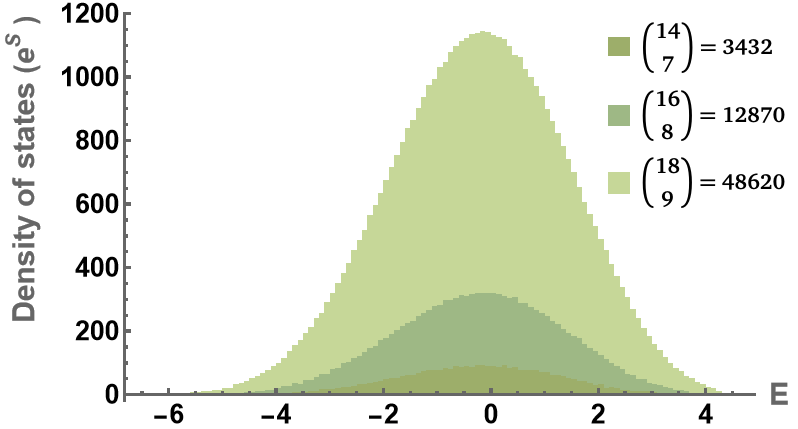}
\caption{\small{Density of states for the different system sizes $L = 14, N=7$, and $L = 16, N=8$, and $L = 18, N=9$ of the chaotic spin chain.}}
\label{expS_comp1}
\end{center}
\end{figure}

In this section, we study the entropic factor $e^{S(\bar{E}+\omega)-S(\bar{E}+\omega/2)}=e^{u_{\bar{E}}(\omega)}$ for finite dimensional state spaces. We consider the chaotic XXZ spin chain with different system sizes $L = 14, N=7$, and $L = 16, N=8$, and $L = 18, N=9$. $L$ is the total number of spin and $N$ is the number of up-spin. The corresponding state spaces have dimensions $\binom{14}{7}=3432$, $\binom{16}{8}=12870$, and $\binom{18}{9}=48620$. Figure \ref{expS_comp1} is the plot of the density of states for the different system sizes under consideration.

\begin{figure}
\begin{center}
\includegraphics[width=.9\linewidth]{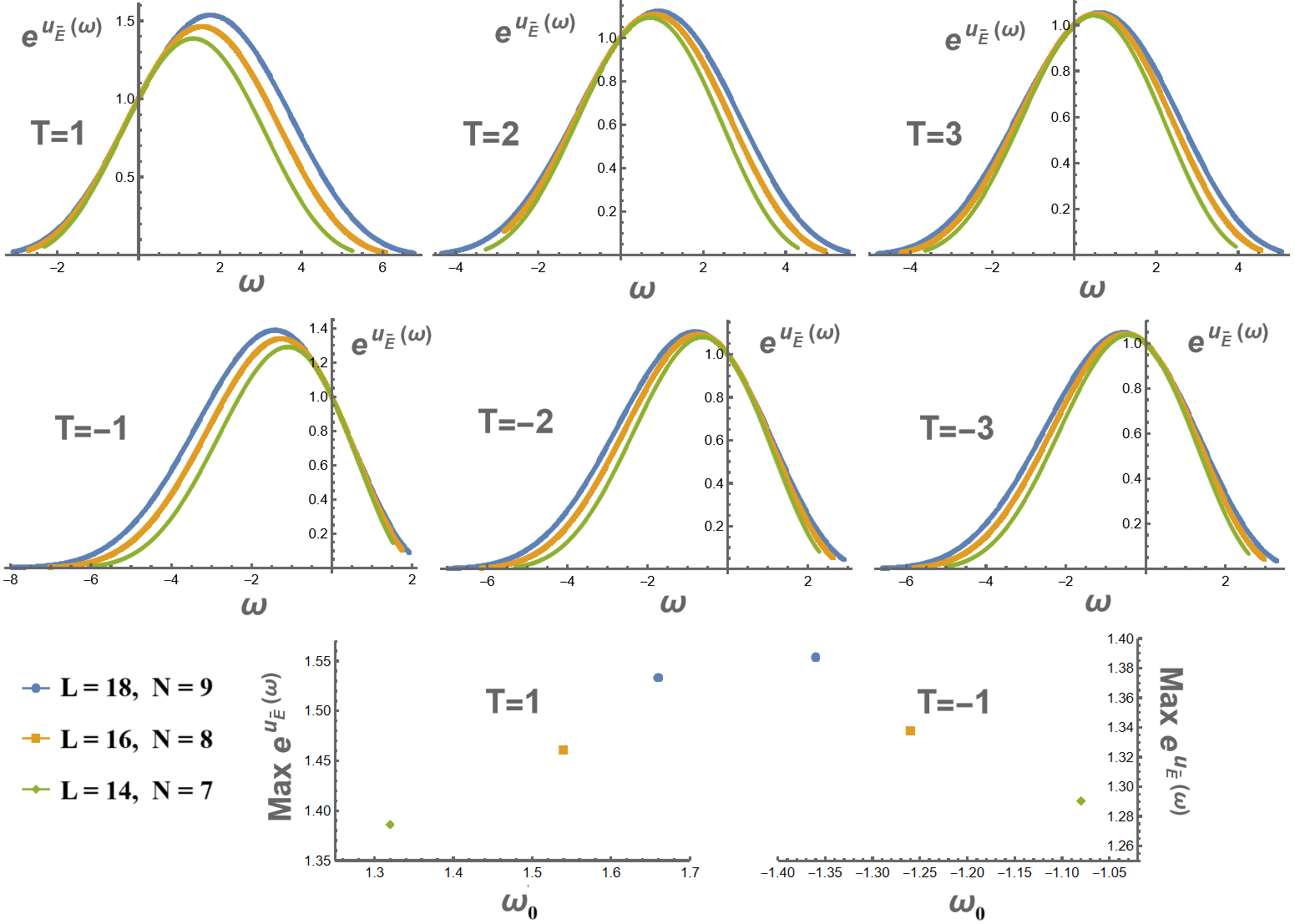}
\caption{\small{Top six subfigures are plots $e^{S(\bar{E}+\omega)-S(\bar{E}+\omega/2)}=e^{u_{\bar{E}}(\omega)}$ as a function of $\omega$ for fixed temperature $T$ ($\bar{E}$ defined in terms of $T$). The different colors are for different system sizes. Bottom two subfigures are plots of Max $e^{u_{\bar{E}}(\omega)}$ as a function of the maxima point $\omega=\omega_0$ for different system sizes at fixed temperatures $T=1$ and $T=-1$. The position of the maxima increases with increasing system size, as we expected from the Taylor expansion of $u_{\bar{E}}(\omega)$.}}
\label{expu_plots}
\end{center}
\end{figure}

We will use canonical entropy for simplicity in this section. So, fixed $E$ corresponds to a fixed temperature $T$. Top six subfigures in Figure \ref{expu_plots} are plots of $e^{u_{\bar{E}}(\omega)}$ for different values of temperature. The plots have maxima at a positive value of $\omega$ for positive temperature and at a negative value of $\omega$ for negative temperature. So for positive temperatures, the entropic factor supports gain in energy. And for negative temperatures, the entropic factor supports lost in energy. This is the contribution of the entropic factor in the arrow of time. For larger systems, the position of the maxima moves away from the origin $\omega=0$. The position of the maxima $\omega=\omega_0$ is the solution of $u_{\bar{E}}'(\omega)=0$. We can get a rough estimate of this position using Taylor expansion about the point $\omega=0$.
\begin{eqnarray}
&& S'(\bar{E}+\omega_0)-S'(\bar{E}+\omega_0/2)=0\nonumber\\
\Rightarrow && -\frac{1}{2}\,\omega_0\frac{\beta^2}{C}+\frac{3}{8}\,\omega_0^2\frac{\beta^2}{C^2}\left(2\beta+\frac{1}{C}\frac{\partial C}{\partial T}\right)\sim 0\nonumber\\
\Rightarrow && \omega_0\sim \frac{4C}{3}\,\frac{1}{\left(2\beta+\frac{1}{C}\frac{\partial C}{\partial T}\right)}\
\end{eqnarray}
where we have used $S''(\bar{E})=-\beta^2/C$ and $S'''(\bar{E})=\beta^2/C^2(2\beta+1/C \partial C/\partial T)$. $C$ is the heat capacity at inverse temperature $\beta$ and it is an extensive quantity. So, $\omega_0$ is an extensive quantity. This agrees with the fact that $\omega_0\to\pm\infty$ in the infinite system size limit when the dimension of the state space also tends to infinity. The maxima of $e^{u_{\bar{E}}(\omega)}$ and its position $\omega_0$ for the different system sizes are plotted in the bottom subfigures of Figure \ref{expu_plots}.

The plots of $e^{u_{\bar{E}}(\omega)}$ also broaden for larger systems. This can also be analysed by Taylor expansion of $u_{\bar{E}}(\omega)$ about the position of the maxima $\omega=\omega_0$.
\begin{eqnarray}
u_{\bar{E}}(\omega)\sim u_{\bar{E}}(\omega_0)-\frac{(\omega-\omega_0)^2}{2C}\,\left(\beta^2_2-\beta^2_1\right)\nonumber\
\end{eqnarray}
where $\beta_1$ and $\beta_2$ are inverse temperatures at $E+\omega_0/2$ and $E+\omega_0$, and again $C$ is the heat capacity of the system. We have ignored the small difference in the heat capacity at inverse temperatures $\beta_1$ and $\beta_2$. $(\beta^2_2-\beta^2_1)$ is not an extensive quantity so again the width of the maxima is controlled by the heat capacity $C$ which is an extensive quantity.

%

\section{Numerical methods and further numerical results}
\label{numres_app}
The main Mathematica notebooks including the codes and the results are available as ancillary files at https://arxiv.org/abs/2212.03914. Around 128 GB of computer memory is required to handle the largest system size under consideration $L=18, N=9$. The ancillary files also include the sorted energy eigenvalues for the three system sizes. So, at least Figures \ref{cmc_comp}, \ref{expS_comp1} and \ref{expu_plots} can be reproduced readily in any computer with bare minimum memory.

The construction of the Hamiltonian and the other operators are straight-forward in the site basis \cite{Santos_lecture_2014}. After solving for the energy eigenvalues and the energy eigenstates, it is again straight-forward to calculate the matrix elements $\mathcal{O}_{mn}$.

For the calculation of $\Delta E_n$ in Figure \ref{6panels_l2}, it is more convenient to simply perform matrix algebra
\begin{equation}
\mathcal{O}H_0\mathcal{O}-\frac{1}{2}\left(\mathcal{O}^2H_0-H_0\mathcal{O}^2\right)
\end{equation}
and calculate the expectation values in the energy eigenstates. Even for the calculation of the expectation values, matrix multiplications with the matrix constructed out of the eigenvectors is parallelized and much faster compared to individual calculation of the expectation value involving the row eigenvector and the column eigenvector. Similarly, we calculate the $\Delta E$ in Figure \ref{l4-l6_XXZ} using the appropriate matrix algebra.

The $f$-function as a function of $\bar{E}$ are calculated for different fixed values of $\omega$. For a given $\omega$, we pick out the matrix elements lying within a small window $d\omega$. We choose $d\omega=0.01$ for the system size $L=14, N=7$; $d\omega=0.005$ for the system size $L=16, N=8$; and $d\omega=0.0005$ for the system size $L=18, N=9$. The number of elements picked out in this process is around $22000$ to $45000$ for the system size $L=14, N=7$; $300000$ to $160000$ for the system size $L=16, N=8$; and $230000$ to $400000$ for the system size $L=18, N=9$. So, even though we are decreasing $d\omega$ superlinearly, the number of elements grows very quickly. After this, the elements are sorted by increasing order of $\bar{E}$. We calculate the average of $|\mathcal{O}_{mn}|^2$ over the window $dE=0.8$. The entropic factor $e^{-S(\bar{E})}$ is cancelled out by multiplying with $e^{S(\bar{E})}$. We used the microcanonical entropy. This gives us $f(\bar{E},\omega)^2$ so we take the square root to get $f(\bar{E},\omega)$.

\begin{figure}
\begin{center}
\includegraphics[width=.8\linewidth]{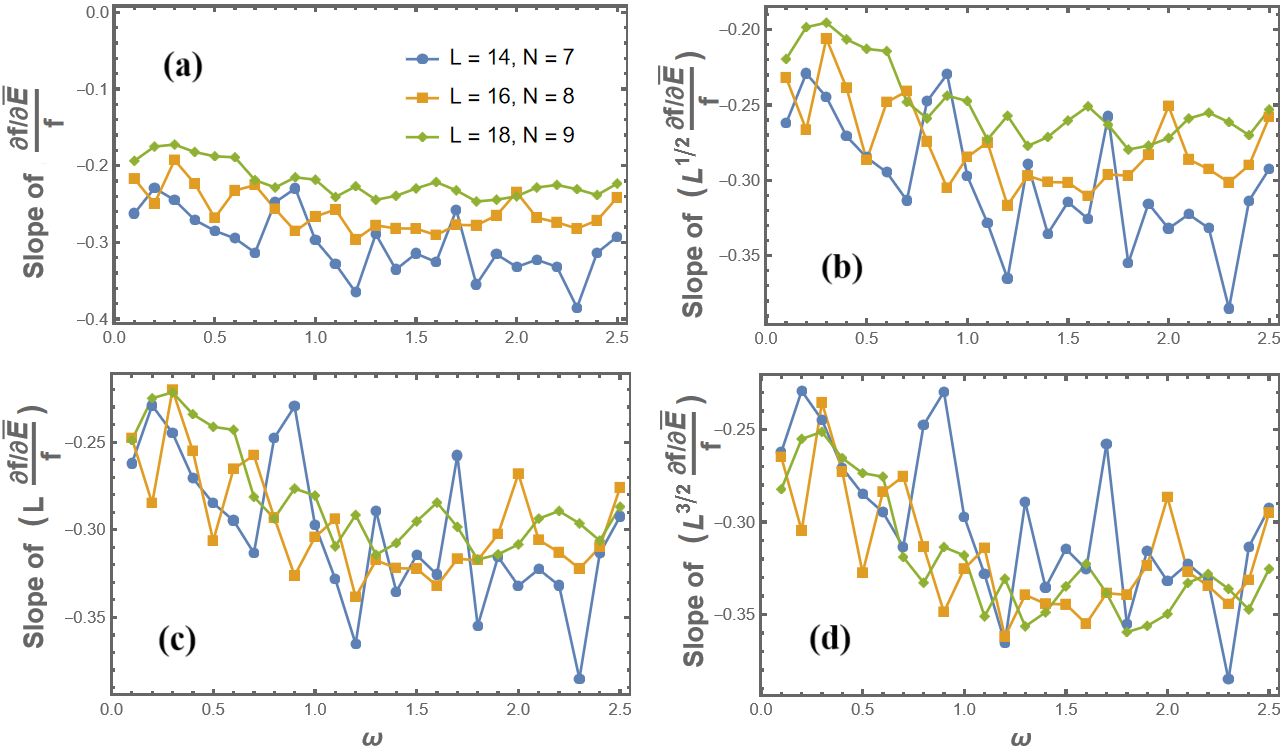}
\caption{\small{Panel (a) is plots of slope of linear fit of $1/f \partial f/\partial \bar{E}$ for $|\bar{E}|\leq 1$ for the kinetic energy operator with different system sizes $L=14, N=7$, $L=16, N=8$, and $L=18, N=9$. Panels (b), (c) and (d) are plots of the slopes after scaling with different powers of the system size. Panel (c) shows that the slope is decreasing as $L^{-1}$ where $L$ is the system size.}}
\label{slopes_dfdEb_ke_plot}
\end{center}
\end{figure}

Panel (a) in Figure \ref{slopes_dfdEb_ke_plot} are the plots of the slope of of the linear fit of $1/f \partial f/\partial \bar{E}$ as a function of $\omega$ for the kinetic energy operator $\mathcal{O}_{KE}$ for different system sizes. Panel (b), (c) and (d) are similar plots but scaled with different system sizes. Panel (c) again shows that $f(\bar{E})$ flattens as $L^1$ as the system size $L$ increases.

\begin{figure}
\begin{center}
\includegraphics[width=.8\linewidth]{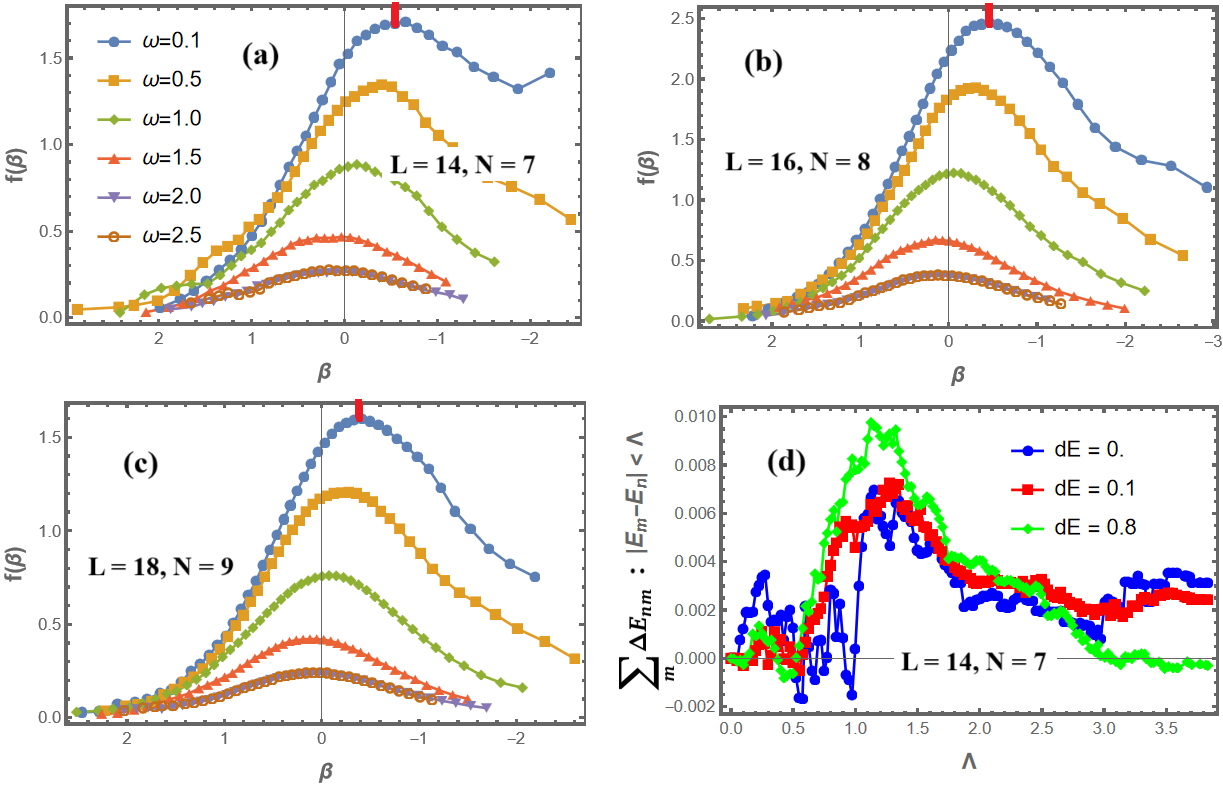}
\caption{\small{Panels (a), (b) and (c) are plots of $f(\beta)$ plots of the kinetic energy operator $\mathcal{O}_{KE}$ with different system sizes $L=14, N=7$ in (a), $L=16, N=8$ in (b), and $L=18, N=9$ in (c). Panel (d) is plots of change in energy $\Delta E$ as a function of $\Lambda$ which is the upper cut-off of $\omega$, starting from states with $\beta=0$ but different $dE=0, 0.1, 0.8$, for perturbation with the kinetic energy operator in the system size $L=14, N=7$.}}
\label{fEb_beta_DEL14_keL14L16L18_plot}
\end{center}
\end{figure}

Panel (a), (b) and (c) in Figure \ref{fEb_beta_DEL14_keL14L16L18_plot} are plots of $f(\beta)$ of the kinetic energy operator for the different system sizes $L=14, N=7$; $L=16, N=8$ and $L=18, N=9$ respectively. The plots support the monotonic behaviour of $f(\beta)$ function. But we can observe something peculiar, the peak of the $\omega=0.1$ plots does not fall on $\beta=0$. Similar departure of the position of the peak from $\beta=0$ is also observed for $\omega=0.5$. We believe that this is a finite size effect. We can see that the peak position, marked by the red tabs on the top $\beta$-axis frames, moves towards $\beta=0$ as the system size increases. This departure of the peak from $\beta=0$ especially in panel (a) suggests that it would be easy to violate the second law of thermodynamics in the smallest system size $L=14, N=7$ by perturbing with the kinetic energy operator with a cut-off of $\omega$ taken to be $\Lambda$. With such a perturbation, the system would still gain energy starting from a state with effective inverse temperature $\beta=0$. This would be violation of the Kelvin statement. We check if this is the case. Panel (d) is plots of the change in energy as a function of the cut-off $\Lambda$. We considered three initial states - an eigenstate, a linear superposition of eigenstates with energy window $dE=0.1$, and a linear superposition of eigenstates with the microcanonical energy window $dE=0.8$. We find that all of these states generally gain energy from the perturbation.

\begin{figure}
\begin{center}
\includegraphics[width=.8\linewidth]{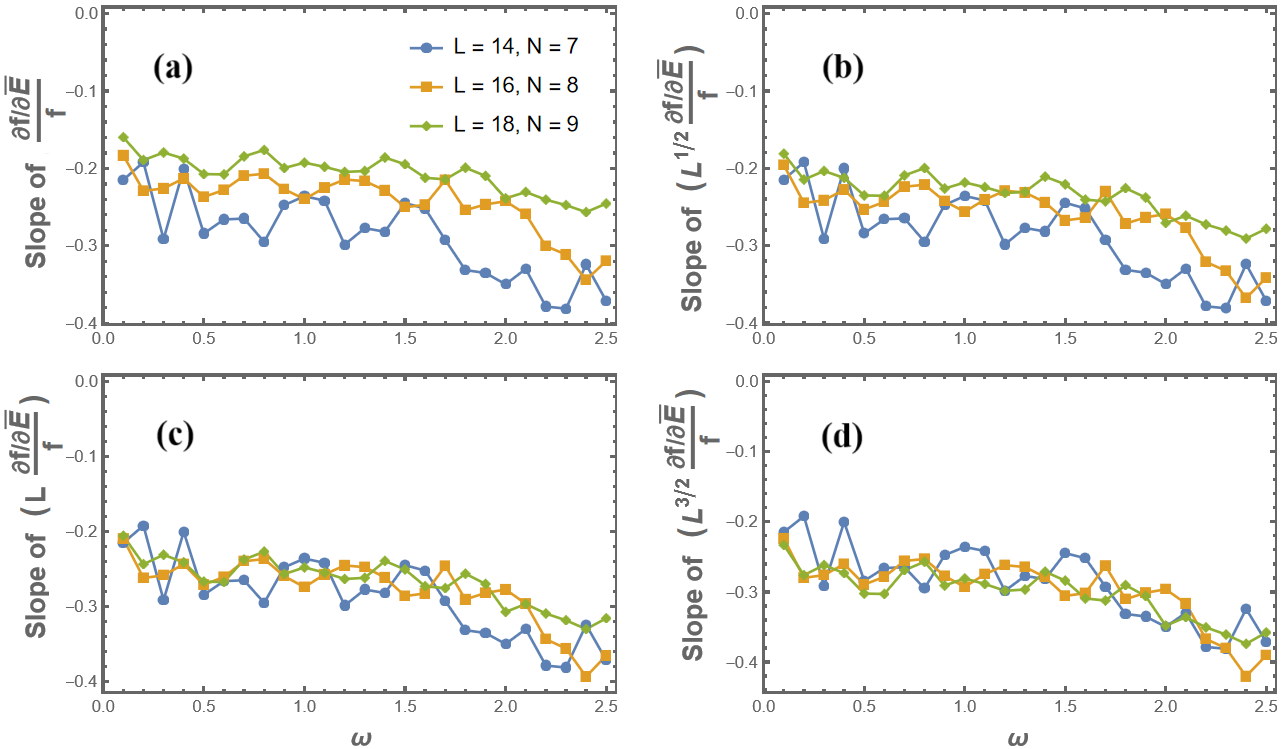}
\caption{\small{Panel (a) is plots of slope of linear fit of $1/f\partial f/\partial \bar{E}$ for $|\bar{E}|\leq 1$ for the single site spin operator $S^z_{L/2}$ with different system sizes $L=14, N=7$ ($7$th site spin operator), $L=16, N=8$ ($8$th site spin operator), and $L=18, N=9$ ($9$th site spin operator). Panels (b), (c) and (d) are plots of the slopes after scaling with different powers of the system size. Panel (c) shows that the slope is decreasing as $L^{-1}$ where $L$ is the system size.}}
\label{slopes_dfdEb_n789_plot}
\end{center}
\end{figure}

We do not consider perturbing the system with the single site spin operator $S^z_{L/2}$. It is because this operator couples state subspaces with different magnetization. But we studied its ETH-monotonic behaviour within the state subspace that we have been working with. We check for the flattening of the $f(\bar{E})$ of the single site spin operator as the system size increases. Panel (a) in Figure \ref{slopes_dfdEb_n789_plot} is the plots of the slope of linear fit of $1/f \partial f/\partial \bar{E}$ as a function of $\omega$ for the single site spin operator $S^z_{L/2}$ for different system sizes. Panel (b), (c) and (d) are similar plots but scaled with different system sizes. Panel (c) again shows that $f(\bar{E})$ flattens as $L^1$ as the system size $L$ increases.

\begin{figure}
\begin{center}
\includegraphics[width=.8\linewidth]{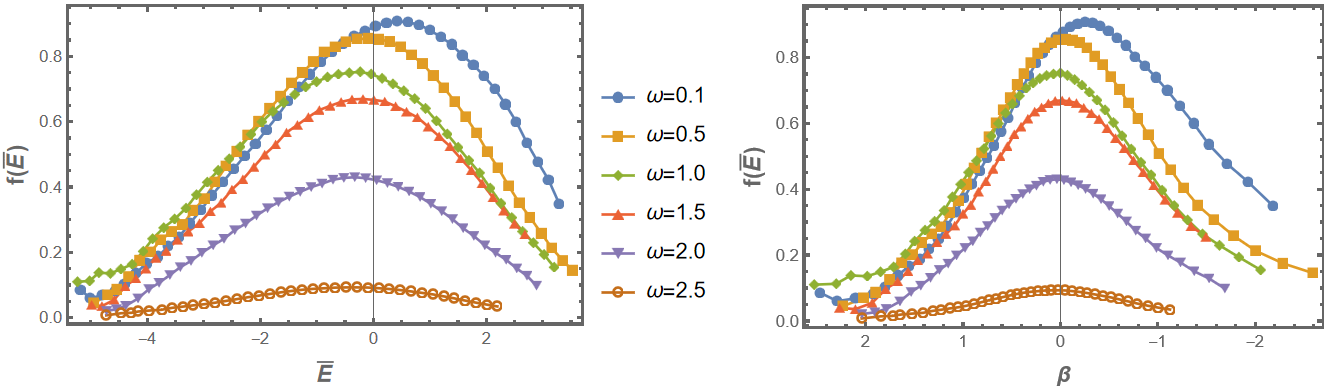}
\caption{\small{Left panel: $f(\bar{E})$ plots with different fixed $\omega$ of the single site spin operator $S^z_{L/2}$ with system size $L=18, N=9$. Right panel: $f(\beta)$ plots with different fixed $\omega$ as a function of inverse effective temperature $\beta$ of the single site spin operator $S^z_{L/2}$ with system size $L=18, N=9$.}}
\label{fEb_Ebbeta_n9_plot}
\end{center}
\end{figure}

Figure \ref{fEb_Ebbeta_n9_plot} are plots of $f(\bar{E})$ and $f(\beta)$ with different fixed $\omega$ of the single site spin operator $S^z_{L/2}$ with system size $L=18$ and $N=9$. We again find that $f(T)$ is a monotonically increasing function of $|T|$.

\bibliography{dE_theorem}
\bibliographystyle{JHEP}

\end{document}